\documentclass[final,twocolumn,10pt,letter]{elsarticle}
\usepackage[a4paper,
    left=1.7cm,
    right=1.7cm,
    top=2.2cm,
    bottom=2.2cm
]{geometry}

\ifdefined\pdfoutput\pdfoutput=1\fi
\newif\ifarchiveversion
\makeatletter
\@ifclassloaded{elsarticle_arxiv}{\archiveversiontrue}{\archiveversionfalse}
\makeatother

\usepackage[utf8]{inputenc}
\usepackage{amsmath,amssymb,amsfonts,amsthm,mathtools}
\usepackage{bm,booktabs,mathrsfs}
\usepackage{array,tabularx}
\usepackage{xcolor}
\usepackage{graphicx}
\usepackage{microtype}
\usepackage[colorlinks=true,linkcolor=blue,citecolor=blue,urlcolor=blue]{hyperref}

\theoremstyle{plain}

\newtheorem{proposition}{Proposition}
\newtheorem{corollary}{Corollary}
\newtheorem{lemma}{Lemma}
\theoremstyle{definition}
\newtheorem{definition}{Definition}
\theoremstyle{remark}
\newtheorem{remark}{Remark}

\newcommand{\R}{\mathbb{R}}
\newcommand{\X}{\mathcal{X}}
\newcommand{\U}{\mathcal{U}}
\newcommand{\Y}{\mathcal{Y}}
\newcommand{\Kop}{\mathcal{K}}
\newcommand{\Gen}{\mathcal{L}}
\newcommand{\WoT}{W_{\mathrm{o},T}}

\newcommand{\rank}{\operatorname{rank}}

\newcommand{\col}{\operatorname{col}}
\newcommand{\ad}{\operatorname{ad}}

\newif\ifdraftnotes
\draftnotestrue

\newcommand{\norm}[1]{\left\lVert#1\right\rVert}
\newcommand{\Lpred}{\mathcal L_{\mathrm{pred}}}
\newcommand{\Lcons}{\mathcal L_{\mathrm{cons}}}

\allowdisplaybreaks

\journal{European Journal of Control}
\begin{document}
\begin{frontmatter}
\title{\vspace{-5mm}%
\LARGE {\bf Learning Control-Relevant Koopman Representations from Input-Output Trajectories}}
\author[aff1]{Maria Bartzioka\corref{cor1}}
\ead{m.bartzioka@tudelft.nl}
\author[aff1]{Mohammad Khosravi}
\cortext[cor1]{Corresponding author}
\address[aff1]{Delft Center for Systems and Control, Delft University of Technology}

\begin{abstract}
Data-driven Koopman representations provide finite-dimensional linear
surrogates for nonlinear dynamics, but predictive accuracy alone does not
ensure suitability for output-feedback control. The lifted state must be
recoverable from measured outputs, the model must adequately represent the
effect of the applied input, and the data must be sufficiently informative
for identification. This paper develops a continuous-time framework for
learning control-relevant Koopman representations using only measured
input-output trajectories. A causal encoder constructs the lifted state
from a finite input-output window, and an additive lifted model is
identified by integral regression. We distinguish observability, input-map
compatibility, controllability, and data informativity. We characterize
compatibility with a constant lifted input matrix, quantify input-map
mismatch, and, under exact lifting assumptions, relate iterated Lie
brackets of the nonlinear system to the controllable subspace of the
lifted realization. A covariance-based normalization controls the scale of
the finite-horizon observability criterion, which is optimized subject to
separate bounds on output-prediction and temporal-consistency errors.
Under the stated exactness and observability assumptions, the encoded
lifted state depends only on the current nonlinear state. For approximate
representations, we bound the difference between encodings of distinct
past trajectories ending at the same state. An exactly liftable controlled
system provides analytical ground truth, while a controlled Duffing
oscillator illustrates the approximate setting, for which an informative
exact additive lifting on an open set is shown to be impossible. On the
exact benchmark, encoder initialization determines whether the learned
representation recovers the analytical lift. In the reported Duffing
experiments, observability promotion improves nominal Gramian measures
without improving observer-based estimation or tracking, while updating
the lifted-state estimate through the encoder at every sampling instant
has a larger effect on tracking accuracy.
\end{abstract}

\end{frontmatter}
\ifarchiveversion\onehalfspacing\fi

\newpage
\section{Introduction}
\label{sec:introduction}

Koopman operator theory provides a framework for describing nonlinear dynamics through the linear evolution of functions of the state \citep{koopman1931hamiltonian,mezic2005spectral,mauroy2020koopman,brunton2022modern}. For an autonomous nonlinear system, observables spanning a finite-dimensional subspace invariant under the Koopman generator evolve according to a linear differential equation \citep{klus2020generator}. Dynamic mode decomposition and its extensions approximate finite-dimensional Koopman representations from data \citep{schmid2010dynamic,williams2015data}. This perspective is attractive for control because finite-dimensional lifted models can retain nonlinear features of the original dynamics while remaining compatible with linear estimation and model predictive control (MPC) \citep{korda2018linear,peitz2020generators,bevanda2021koopman}. Controlled extensions of Koopman and dynamic-mode-decomposition methods incorporate measured inputs into lifted predictors \citep{proctor2016dmdc,proctor2018inputs,williams2016extending,korda2018linear}; moreover the observables defining the representation can also be learned from data rather than fixed in advance \citep{takeishi2017learning,lusch2018deep,otto2019linearly,korda2020optimal,khosravi2021learning,khosravi2023representer,haseli2022learning}. This flexibility raises the question of which properties the learned representation should possess for output-feedback control.

Predictive accuracy alone does not resolve this question. When the full nonlinear state is not measured, the controller must infer the lifted state from the available outputs. For example, deep subspace encoders use past inputs and outputs
to initialize models trained with a truncated prediction loss \citep{beintema2023deep}.
However, a representation may predict measured trajectories accurately once its initial lifted state is supplied and still be poorly suited to output feedback if distinct lifted states
generate only weakly distinguishable output trajectories. This raises questions about the observability of the learned representation and the sensitivity of lifted-state recovery to measurement noise. Observability of Koopman representations has been studied
from nonlinear and output-based viewpoints
\citep{mesbahi2021nonlinear,park2024observability,otto2021koopman}. Koopman-based observers exploit lifted dynamics for state estimation \citep{surana2016linear,surana2016koopman}. Related Kazantzis--Kravaris/Lu\-en\-ber\-ger-type observers seek transformed state variables with linear dynamics driven by the measured output \citep{andrieu2006existence,janny2021deep}. For autonomous systems, Takens' embedding theorem establishes that, under suitable genericity assumptions, delay coordinates formed from a scalar observable can provide an embedding of the underlying state manifold \citep{takens1981detecting}. This motivates the use of delayed measurements in Koopman-based representations \citep{arbabi2017ergodic,brunton2017chaos,arbabi2018datadriven,kamb2020timedelay}. For controlled systems, the effect of the applied inputs must also be accounted for when constructing a representation from measured trajectories \citep{otto2024bilinear}. These considerations motivate learning representations whose lifted states can be reliably distinguished from measured output trajectories. For a fixed normalization, weak output sensitivity to changes in the lifted state makes its recovery sensitive to measurement noise. We therefore optimize a regularized log-determinant criterion of the finite-horizon observability Gramian after covariance-based normalization, subject to separate bounds on output-prediction and
temporal-consistency errors. The latter measure agreement between model-propagated lifted states and encodings of the corresponding later input-output windows. Whether improvements in the nominal observability measure translate into better estimation and control is assessed separately.

Neural encoders that map past input-output measurements to latent states have been used in nonlinear system identification and Koopman learning \citep{iacob2021deep,beintema2023deep,masti2021learning,iacob2025noise}. In the framework proposed in the current paper, a causal encoder constructs the lifted state from a finite input-output window. For the observer-based controller, this encoding initializes the lifted-state estimate, which is subsequently propagated and corrected using the learned model and incoming measurements. A representation constructed from a trajectory window need not depend only on the current nonlinear state. We first establish sufficient conditions under which an exact lifted model assigns the same encoded lifted state to distinct past trajectories ending at the same nonlinear state. We then derive a bound on the difference between these encodings in terms of their
output-prediction errors and the smallest eigenvalue of the observability Gramian. This second result addresses the approximate
setting relevant to learned models.

Recovering the lifted state from measured input-output trajectories addresses the estimation component of output-feedback control. Predicting the effect of future control actions additionally requires an appropriate representation of the input in the lifted dynamics. First, for a control-affine nonlinear system, the effect of the input on the lifted dynamics is generally state dependent. An additive linear surrogate instead represents this effect by a constant lifted input matrix. State-dependent and bilinear Koopman formulations represent this dependence more explicitly \citep{goswami2017bilinear,bruder2021bilinear,folkestad2021koopman,kieboom2023utility,iacob2024koopman,otto2024bilinear}. An additive model with a constant input matrix therefore imposes an additional structural restriction, whose compatibility with the chosen representation is examined in Section~\ref{sec:properties}. Second, controllability of the learned realization $(A,B)$ and informativity of the data used to identify $A$ and $B$ are different properties. The former concerns how the input acts on the learned realization, whereas the latter concerns whether the data contain sufficient independent excitation to identify its autonomous and forced components \citep{willems2005pe,gevers2009identification,vanwaarde2020data}, which can be compromised under feedback without sufficient independent excitation \citep{vandenhof1995identification,forssell1999closedloop}.

\citet{park2024observability} study observability of Koopman representations and develop an output-based identification method using an observable canonical construction. \citet{strasser2026io} construct a bilinear surrogate from input-output trajectories and develop an output-feedback controller with closed-loop stability guarantees under their stated assumptions. These studies address structural observability and guaranteed
output-feedback control, respectively. For the additive models considered here, we examine the interplay between finite-horizon output sensitivity, prediction accuracy, and compatibility with a constant lifted input matrix. Our focus is on learning an additive continuous-time representation by optimizing a regularized log-determinant criterion of its finite-horizon observability Gramian, after covariance-based
normalization and subject to separate prediction and consistency bounds. We also examine the structural restrictions of this model class.

This formulation builds on our earlier observability-aware learning framework for autonomous discrete-time systems \citep{bartzioka2026maximally}, which combined an observability criterion with invariance, output-fit, and reconstruction terms in a weighted loss. The present work extends this framework to controlled, non-autonomous, continuous-time systems using measured input-output trajectories and promotes observability subject to separate bounds on output-prediction and temporal-consistency errors. A nonlinear full-state reconstruction map is not required for the proposed output-feedback implementation, which uses lifted-state estimates and the linear output map $C$. We use a continuous-time formulation to express input-map compatibility and the relation to nonlinear Lie brackets directly in terms of the system's vector fields. Integral regression
identifies the corresponding model matrices without constructing numerical derivative targets. For sampled estimation and predictive control, the learned nominal model is discretized under
zero-order-held inputs as in~\eqref{eq:zoh}. A directly identified discrete-time predictor is an alternative, but the structural conditions established here concern the continuous-time model and cannot be transferred to that predictor without a separate analysis.

The main contributions are:
\begin{enumerate}
    \item A continuous-time formulation for learning observability-aware
    Koopman representations from measured input-output trajectories, in
    which finite-horizon observability is promoted subject to separate
    prescribed bounds on output-prediction and temporal-consistency
    errors. This formulation makes the trade-off between observability
    and predictive accuracy explicit. Covariance-based normalization
    controls the scale of the observability criterion, while a nominal
    Gaussian estimation model relates the objective to uncertainty in
    lifted-state recovery (Proposition~\ref{prop:posterior}).

    \item An analysis connecting trajectory-based encoders to state-based
    Koopman liftings and clarifying the structural restrictions of
    additive lifted models. Under the stated exactness, observability,
    and trajectory assumptions, the encoded lifted state depends only
    on the current nonlinear state
    (Proposition~\ref{prop:encoder_factorization}). For approximate
    representations, prediction errors and the smallest observability
    Gramian eigenvalue bound the difference between encodings of distinct
    past trajectories ending at the same state
    (Proposition~\ref{prop:code_disagreement}). We also characterize 
    compatibility with a constant lifted input matrix and, under exact 
    lifting assumptions, relate iterated Lie brackets of the nonlinear 
    system to the lifted controllable subspace (Proposition~\ref{prop:bracket_krylov}). 
    The resulting rank bound (Corollary~\ref{cor:controllability_limitation}) shows that 
    an exact additive lift with more coordinates than the original nonlinear state cannot
    be fully controllable.

    \item An identification and output-feedback control framework that
    combines the learned representation with integral regression,
    lifted-state estimation, and model predictive control, while
    distinguishing data informativity from controllability. Evaluation
    on an exactly liftable controlled system and a controlled Duffing
    oscillator examines the practical implications of the analysis.
    The numerical results distinguish improvements in nominal
    observability from improvements in estimation and control:
    in the reported Duffing experiments, improved Gramian measures
    do not yield improved observer-based estimation or tracking.
\end{enumerate}

The remainder of the paper is organized as follows. Section~\ref{sec:problem_formulation} introduces the nonlinear system, Koopman lifting, and output-based recovery of lifted states. Section~\ref{sec:properties} studies the control-relevant properties of the additive lifted realization. Section~\ref{sec:learning} presents the learning and identification procedure and the output-feedback controller. Section~\ref{sec:examples} presents validation on an exactly liftable controlled system and the controlled Duffing study. Section~\ref{sec:conclusions} concludes the paper.

\section{Problem formulation and preliminaries}
\label{sec:problem_formulation}
\label{sec:background}
\subsection{Problem formulation}
\label{subsec:problem_setting}

Consider the continuous-time nonlinear control-affine system
\begin{equation}
    \dot{x}(t)
    =
    f\bigl(x(t)\bigr)
    +
    g\bigl(x(t)\bigr)u(t),
    \;
    y(t)=h\bigl(x(t)\bigr),
    \label{eq:nonlinear_ct}
\end{equation}
where $x(t)\in\X\subseteq\R^n$ is the state,
$u(t)\in\U\subseteq\R^m$ is the applied control input, and
$y(t)\in\Y\subseteq\R^p$ is the ideal system output, before measurement
noise is added. The maps
$f:\X\rightarrow\R^n$,
$g:\X\rightarrow\R^{n\times m}$, and
$h:\X\rightarrow\R^p$ are assumed continuously differentiable, with any additional smoothness required for the higher-order Lie-bracket analysis stated explicitly in Section~\ref{sec:properties}. For multiple inputs, write $   g(x)
    =
    \begin{bmatrix}
        g_1(x) & \cdots & g_m(x)
    \end{bmatrix},
    \;
    g_j:\X\rightarrow\R^n $, so that $g(x)u=\sum_{j=1}^{m}g_j(x)u_j$. Here, $f(x)$ describes the autonomous dynamics of the system, whereas each $g_j(x)$ specifies how the corresponding control input $u_j$ acts on the state. This decomposition allows the autonomous and input-dependent components of the lifted dynamics to be examined separately. At the generator level, the corresponding separation into an autonomous term and terms proportional to the inputs underlies controlled Koopman formulations such as those in \citet{peitz2020generators,otto2024bilinear}; it does not by itself imply a constant additive input matrix in a finite-dimensional lifted model.

The maps are defined on an open neighbourhood of the operating region,
and trajectories are considered only while they remain in that region.
Inputs are piecewise continuous. The admissible input set contains a neighbourhood of zero, allowing the
autonomous and input-dependent parts to be separated. For the MPC formulation,
$\U$ is a compact convex polytope, representing bounded actuator constraints
through finitely many linear inequalities. This choice preserves the
convexity of the nominal linear MPC problem with a convex quadratic cost.
At sampling times $t_k=k\Delta t$,
measurements are $y_k^{\rm m}=h(x(t_k))+v_k$, where $v_k$ denotes
measurement noise.

The full state $x(t)$ is not assumed to be available. Instead, the applied
input $u(t)$ is known and measurements of the output $y(t)$ are available. The output map
$h$ is not assumed to be injective; hence, distinct states may produce the same instantaneous output, $x_1\neq x_2$ while $h(x_1)=h(x_2)$. Consequently, a single measurement $y(t)$ need not uniquely determine the state $x(t)$. This does not imply that the system is unobservable, since measurements collected over time may distinguish states that cannot be separated from a single output sample. The objective is therefore to construct a finite-dimensional lifted
representation from the available input-output information, $z^{\mathrm{e}}(t)
    =
    \Phi_\theta\bigl(\eta(t)\bigr)
    \in\R^N$, where $\Phi_\theta$ denotes a parameterized encoder and $\eta(t)$ denotes a finite causal segment of measured outputs and known inputs, whose precise construction is introduced in
Section~\ref{subsec:output_recovery}. We seek a representation whose
controlled evolution is described by the finite-dimensional surrogate
\begin{equation}
    \dot z(t)
    \approx
    Az(t)+Bu(t),
    \;
    \hat y(t):=Cz(t),
    \label{eq:problem_target_model_ct}
\end{equation}
with $A\in\R^{N\times N}$, $B\in\R^{N\times m}$, and
$C\in\R^{p\times N}$.

The control task is constrained tracking of feasible piecewise constant
output references, including regulation to an equilibrium. We express
the state, input, and output relative to a known equilibrium pair
$(x_\star,u_\star)$ satisfying
$f(x_\star)+g(x_\star)u_\star=0$, with $y_\star=h(x_\star)$.
Specifically, the deviations are $\tilde x=x-x_\star$,
$\tilde u=u-u_\star$, and $\tilde y=y-y_\star$, with the admissible
input set translated accordingly. The transformed autonomous dynamics
are $\tilde f(\tilde x)=f(x_\star+\tilde x)+g(x_\star+\tilde x)u_\star$,
and the transformed output map is
$\tilde h(\tilde x)=h(x_\star+\tilde x)-y_\star$.
Dropping the deviation notation gives $f(0)=0$ and $h(0)=0$;
this coordinate convention does not imply that the equilibrium is stable.
The present framework evaluates representation
quality and output-feedback performance. Asymptotic regulation, offset-free
tracking and recursive feasibility require additional conditions and are not
inferred from a small fitting error. Beyond reproducing measured trajectories, the learned representation should retain information that can be inferred from the measured outputs and should capture the effect of the applied control input sufficiently well for output-feedback control. These requirements motivate the two developments that follow: Section~\ref{subsec:koopman_lifting}
introduces the Koopman framework underlying the lifted dynamics in
\eqref{eq:problem_target_model_ct}, while
Section~\ref{subsec:output_recovery} introduces the observability concepts
and input-output trajectory information used to recover the lifted state.

\subsection{Koopman lifting of the nonlinear dynamics}
\label{subsec:koopman_lifting}

We first consider the autonomous dynamics
\begin{equation}
    \dot x=f(x),
\end{equation}
and denote its flow map by $\varphi^t$, so that $\varphi^t(x_0)$ is
the state reached at time $t$ from the initial condition $x_0$,
for times at which the trajectory remains in the domain under consideration.
For a scalar observable $\psi:\X\to\R$, the Koopman semigroup $\{\Kop^t\}_{t\ge0}$ acts by composition with the flow,
\begin{equation}
(\Kop^t\psi)(x)
    :=
    \psi\bigl(\varphi^t(x)\bigr).
\end{equation}
Although the state dynamics are nonlinear, each operator $\Kop^t$ is linear on the space of observables \citep{koopman1931hamiltonian,brunton2022modern}. The corresponding infinitesimal generator is $(\Gen\psi)(x)
    :=
    \lim_{t\downarrow0}
    \frac{(\Kop^t\psi)(x)-\psi(x)}{t}
    =
    \nabla\psi(x)^{\top}f(x)$, whenever the derivative exists, where $\nabla\psi$ denotes the gradient of the scalar observable \citep{klus2020generator}. The generator is particularly useful here because it gives a differential description of the observable dynamics and therefore matches the continuous-time models considered throughout the paper.

Let $\psi_1,\ldots,\psi_N$ be a finite collection of scalar observables and define the vector-valued lifting
\begin{equation}
    \Psi(x)
    :=
    \begin{bmatrix}
        \psi_1(x)&\cdots&\psi_N(x)
    \end{bmatrix}^{\top},
    \label{eq:finite_lift}
\end{equation}
where $z=\Psi(x)\in\R^N$. If $\operatorname{span}\{\psi_1,\ldots,\psi_N\}$ is invariant under the generator, there exists $A\in\R^{N\times N}$ such that
\begin{equation}
    \dot z=Az.
\end{equation}
When the measured output lies in the span of the selected observables, it can be represented exactly by a linear map
\begin{equation}
    y = Cz,
    \;
    C\in\mathbb{R}^{p\times N}.
\end{equation}
When such an exact representation is not available, as is generally the case for an approximate or learned finite-dimensional representation, we instead impose the linear readout $\hat y := Cz$, with $C$ learned so that $\hat y$ approximates the measured output $y$.
Retaining a linear output map gives the pair $(C,A)$ a direct
systems-theoretic interpretation. The existence of the infinite-dimensional Koopman operator does not imply the existence of a finite-dimensional invariant subspace that also retains the state or output information required for control. Exact finite-dimensional closure is exceptional for general nonlinear systems, particularly when the selected invariant subspace is required to contain the original state \citep{brunton2016koopman}. So, we seek an approximately invariant representation over the region of the state space represented by the available data, $\dot z\approx Az,\;\hat y:=Cz$, rather than a global exact linearization. Motivated by this autonomous construction, for the controlled system \eqref{eq:nonlinear_ct} we adopt the additive lifted surrogate already introduced in \eqref{eq:problem_target_model_ct}. The additive form is attractive
because it supports linear estimation and MPC in the lifted space
\citep{proctor2016dmdc,proctor2018inputs,korda2018linear}. Whether a constant matrix $B$ can faithfully represent the nonlinear input action is a separate structural question and it is treated in Section~\ref{subsec:additive_compatibility}. Before addressing that question, we consider how the lifted state can be recovered when full-state measurements are unavailable.

\subsection{Output-based recovery of the lifted states}
\label{sec:controlled_lifting}
\label{subsec:output_recovery}

For the autonomous lifted model
\begin{equation}
    \dot z=Az,
    \;
    \hat y:=Cz,
\end{equation}
observability characterizes whether different lifted initial states can be distinguished from their output trajectories. Since state recovery and control are performed over finite time intervals, we use the finite-horizon observability Gramian. This Gramian is well-defined for any finite horizon $T>0$ without requiring $A$ to be Hurwitz, i.e., without requiring all its eigenvalues to have strictly negative real parts:
\begin{equation}
    W_{\mathrm{o},T}(A,C)
    :=
    \int_0^T
    e^{A^{\top}t}C^{\top}Ce^{At}\,dt.
    \label{eq:ct_obs_gramian}
\end{equation}
For two initial lifted states $z_0^{(1)}$ and $z_0^{(2)}$, let
$\Delta z_0=z_0^{(1)}-z_0^{(2)}$. Their output difference is
$\hat y^{(1)}(t)-\hat y^{(2)}(t)=Ce^{At}\Delta z_0$.
Integrating its squared Euclidean norm gives the finite-horizon
output separation
\begin{equation}
    \int_0^T
    \|\hat y^{(1)}(t)-\hat y^{(2)}(t)\|_2^2\,dt
    =
    \Delta z_0^{\top}
    W_{\mathrm{o},T}(A,C)
    \Delta z_0.
    \label{eq:ct_output_separation_main}
\end{equation}
Equation \eqref{eq:ct_output_separation_main} gives a direct interpretation of the Gramian. A unit vector $v\in\R^N$ specifies a perturbation of the initial lifted state, whose accumulated squared output response is
$v^\top W_{\mathrm{o},T}(A,C)v
    =
    \int_0^T \|Ce^{At}v\|_2^2\,dt$.
If $v$ is a unit eigenvector of $W_{\mathrm{o},T}$ with eigenvalue
$\lambda$, this quantity equals $\lambda$. Thus, lifted-state
perturbations along eigenvectors associated with small Gramian
eigenvalues produce only weak changes in the output over the horizon
$[0,T]$ and are difficult to distinguish in the presence of measurement
noise. This motivates treating observability as a design objective for the learned representation rather than only as a post-identification diagnostic.

An instantaneous output $y(t)$ need not contain enough information to determine a useful lifted state, whereas a finite segment of past outputs can contain information about the underlying dynamics. For autonomous systems, this use of delayed measurements is motivated by Takens' embedding theorem under its stated genericity assumptions \citep{takens1981detecting}; the theorem does not directly guarantee reconstruction for the controlled, noisy setting considered here. Because those outputs were generated under known control actions, the applied input trajectory is included as well. For delays
$0=\tau_0<\tau_1<\cdots<\tau_{d_y-1}$ and
$0<\tau^u_1<\cdots<\tau^u_{d_u}$, define the finite causal input-output trajectory vector
\begin{equation}
\begin{split}
    \eta(t):=\col\bigl(y(t-\tau_{d_y-1}),&\ldots,y(t),\\
                      u(t-&\tau^u_{d_u}),\ldots,u(t-\tau^u_1)\bigr),
    \end{split}
    \label{eq:history_vector_ct}
\end{equation}
For equally spaced samples, the experimental window is
$\eta_k=\col(y^{\rm m}_{k-L:k},u_{k-L:k-1})$, containing $L+1$ outputs
and $L$ past inputs. The output samples are collected at the endpoints
of $L$ sampling intervals, while the $L$ inputs are applied over those
intervals. The current input $u_k$ is selected after the measurement
$y_k^{\rm m}$ becomes available and is therefore excluded from the
window. In~\eqref{eq:history_vector_ct}, $\col(\cdot)$ denotes vertical stacking, $d_y$ and $d_u$ denote the numbers of output and input samples included in the delay vector, while $\tau_i$ and $\tau^u_i$ denote their corresponding time delays. Only present and past data are used, so the construction is causal. In learning
and online estimation, measured outputs $y^{\rm m}$ replace $y$ in this vector;
exact theoretical statements use noise-free windows. Known past inputs are
included by default: even an additive linear realization has a stacked-output
relation $Y=\mathcal O z_0+\mathcal T U$, so its input-driven response must be
accounted for when recovering the latent state. An output-only window, obtained
by setting $d_u=0$, may be examined as a comparison to assess the effect
of omitting past inputs. We denote it by
\begin{equation}
    \eta_y(t)
    :=
    \col\bigl(y(t-\tau_{d_y-1}),\ldots,y(t-\tau_1),y(t)\bigr),
    \label{eq:eta_y}
\end{equation}
while the input-output vector~\eqref{eq:history_vector_ct} is used by default (Section~\ref{sec:output_history}). The trajectory-window length and delay locations are treated as design choices and selected from validation data relative to the time scales represented in the measurements. The data-driven lifted representation is then constructed as $z^{\mathrm{e}}(t)=\Phi_\theta\bigl(\eta(t)\bigr)$. This encoder must be distinguished from the state-based map $\Psi(x)$ in \eqref{eq:finite_lift}. The latter is used to state structural properties of a Koopman lifting when the nonlinear state is available conceptually, whereas $\Phi_\theta$ constructs the lifted representation from measured input-output information. The superscript $\mathrm{e}$ labels the encoded lifted-state vector $z^{\mathrm{e}}$ and distinguishes it from a state-based lifted-state vector $z=\Psi(x)$ and an observer estimate $\hat z$. The two descriptions coincide when the encoder depends on the input-output window only through the current nonlinear state; equivalently, there exists a state-based map $\Psi_\theta$ such that
\begin{equation}
    \Phi_\theta\bigl(\eta(t)\bigr)
    =
    \Psi_\theta\bigl(x(t)\bigr)
    \label{eq:factorization}
\end{equation}
for the admissible trajectories under consideration. Here, the subscript
$\theta$ indicates that $\Psi_\theta$ is the state-based map induced by
the particular encoder $\Phi_\theta$, when such a map exists; it is not
a separate network trained from full-state measurements. Under this
interpretation, the finite input-output segment provides the information
needed to infer a state-dependent lifted-state vector, and the structural
analysis formulated in terms of $\Psi$ applies, provided the differentiability
and exactness assumptions of the corresponding result also hold. In the general data-driven setting, however, this factorization is not
assumed a priori: $\Phi_\theta$ is learned directly from causal
input-output information and is required to produce a representation that
approximately obeys the controlled lifted dynamics in~\eqref{eq:problem_target_model_ct}.
Proposition~\ref{prop:encoder_factorization} gives sufficient conditions for
exact factorization; Proposition~\ref{prop:code_disagreement} bounds the
remaining dependence on the past under approximate prediction.
\section{Control-relevant properties of Koopman representations}
\label{sec:properties}

The additive realization
\begin{equation}
    \dot z=Az+Bu,
    \qquad
    y=Cz
    \label{eq:exact_additive_lti_for_obs}
\end{equation}
raises three distinct control-relevant questions concerning the same lifted model. 
\begin{enumerate}
    \item \emph{Lifted observability} characterizes whether different lifted states can be distinguished/recovered from the measured output when the input is known. 
    \item \emph{Input-map compatibility} characterizes whether the matrix $B$ faithfully represents the action of the nonlinear input vector fields after lifting.
    \item \emph{Lifted controllability} characterizes the directions of the finite-dimensional realization that can be influenced through this input map.
\end{enumerate}

These questions concern different components of the same model and are treated
as such below. Statements involving the nonlinear system are conditional
on the exact realization or state-based lifting assumptions stated with each
result; these assumptions are not imposed on every learned window predictor.

\textbf{Notation.} $\|\cdot\|_2$ denotes the Euclidean norm for vectors and its induced norm for matrices. For a time-dependent vector, $\|\cdot\|_{L^2(0,T)}$ denotes
the corresponding norm integrated over $[0,T]$.

\subsection{Observability of the lifted realization}
\label{subsec:known_input_observability}

For the additive lifted realization \eqref{eq:exact_additive_lti_for_obs}, the output at time $t_0+t$
resulting from the initial lifted state $z(t_0)$ and the applied input over
$[t_0,t_0+t]$ is
\begin{equation}
    y(t_0+t)
    =
    Ce^{At}z(t_0)
    +
    \int_0^t
    Ce^{A(t-s)}B\,u(t_0+s)\,ds.
    \label{eq:forced_output_ct}
\end{equation}

\begin{lemma}[Output separation under a common known input]
\label{lem:known_input_cancellation}
Let $T>0$ and consider two trajectories of
\eqref{eq:exact_additive_lti_for_obs} on $[t_0,t_0+T]$ with initial
conditions $z^{(1)}(t_0)$ and $z^{(2)}(t_0)$. If both trajectories are
driven by the same known input $u$, then
\begin{equation}
    y^{(1)}(t_0+t)-y^{(2)}(t_0+t)
    =Ce^{At}\bigl(z^{(1)}(t_0)-z^{(2)}(t_0)\bigr),
    \label{eq:ct_known_input_cancel}
\end{equation}
for any $t\in[0,T]$.
Consequently, their finite-horizon output separation, as defined in
\eqref{eq:ct_output_separation_main}, is independent of the common
known input and is determined by $W_{\mathrm{o},T}(A,C)$, as given in \eqref{eq:ct_obs_gramian} and the initial state difference.
\end{lemma}

\begin{proof}
The input-driven response term in \eqref{eq:forced_output_ct} is identical for the two trajectories and therefore cancels upon subtraction. Integrating the squared Euclidean norm of \eqref{eq:ct_known_input_cancel} over $[0,T]$ yields the quadratic form in \eqref{eq:ct_output_separation_main}.
\end{proof}

The use of the same known input isolates the effect of the initial lifted state: differences between the two output trajectories are then due to the initial-state difference rather than to different control actions. Lemma~\ref{lem:known_input_cancellation} therefore justifies using
$W_{\mathrm{o},T}(A,C)$ to assess finite-horizon observability of the
controlled lifted realization. 


\begin{remark}[Direct feedthrough]
If the measured output has the form $y=Cz+Du$, the term $Du$ represents direct feedthrough, meaning that the current input
affects the output instantaneously rather than only through the lifted-state
dynamics. For two trajectories driven by the same known input, this
contribution is identical and therefore cancels in their output difference.
Hence, Lemma~\ref{lem:known_input_cancellation} remains unchanged. For the
remainder of this paper, we assume no direct feedthrough, i.e., $D=0$.
\end{remark}

Let $T_\eta$ be the largest delay in~\eqref{eq:history_vector_ct}, defined as $T_\eta:=\max\bigl(
\{\tau_i:0\le i<d_y\}
\cup\{\tau_j^u:1\le j\le d_u\}\bigr)$. For an open set $\mathcal D\subseteq\X$, an admissible noise-free window at time $t$ is the vector $\eta(t)$ formed according to~\eqref{eq:history_vector_ct} from a trajectory of~\eqref{eq:nonlinear_ct} defined on $[t-T_\eta,t]$, with its state in $\mathcal D$ throughout that interval and its input
in $\mathcal U$. Denote the set of these windows by
$\mathcal H_{\mathcal D}$. A window reaches $x$ if its generating trajectory satisfies $x(t)=x$. The admissible trajectory family is assumed to include local continuations under any admissible piecewise-continuous future input, for as long as the state
remains in $\mathcal D$.

\begin{proposition}[State-based realization of an exact encoder]
\label{prop:encoder_factorization}
Let $r\ge1$ be an integer and assume that $f$ and $h$ are of class $C^r$ on $\mathcal D$. Let $\mathcal O\subseteq\mathcal D$ be an open set, and assume that every $x\in\mathcal O$ is reached by at least one window in $\mathcal H_{\mathcal D}$. Suppose that, along every
admissible trajectory in $\mathcal D$ and its local
continuations, the encoded state
$z^{\mathrm{e}}(t)=\Phi(\eta(t))$ is locally absolutely continuous and satisfies
\[
\begin{aligned}
\dot z^{\mathrm{e}}(t)
&=Az^{\mathrm{e}}(t)+Bu(t),\quad\text{a.e. in }t,\\
h(x(t))&=Cz^{\mathrm{e}}(t)
\end{aligned}
\]
on each time interval for which
$\eta(t)\in\mathcal H_{\mathcal D}$.
If $(C,A)$ is observable, then there exists a unique
$\Psi\in C^r(\mathcal O;\mathbb R^N)$ such that
\[
\Phi(\eta(t))=\Psi(x(t))
\;\forall x(t)\in\mathcal O.
\]
Moreover, we have
\begin{equation}
    \begin{split}
    J_\Psi(x)f(x) &= A\Psi(x),
    \\
    J_\Psi(x)g(x) &= B,
    \\
    h(x) &= C\Psi(x),
    \end{split}
\end{equation}
for all $x\in\mathcal O$.
\end{proposition}

\begin{proof}
Choose two admissible input-output windows reaching the same
state $x_0\in\mathcal O$, and denote their encoded states by
$z_{1,0}=\Phi(\eta_1)$ and $z_{2,0}=\Phi(\eta_2)$. Continue both trajectories with $u=0$. Because $x_0\in\mathcal D$ and $\mathcal{D}$ is an open set, these continuations exist and remain in $\mathcal D$ on some interval $[0,\tau]$, $\tau>0$. In addition, both continuations solve $\dot{x}=f(x),\;x_0=x(0)$. Uniqueness therefore implies
that their future states and outputs coincide. The encoded states satisfy the exact lifted model, so their output difference is:
\[
Ce^{As}(z_{1,0}-z_{2,0})=0,
\qquad\forall s\in[0,\tau].
\]
Writing $\delta z_0=z_{1,0}-z_{2,0}$ gives
\[
0
=
\int_0^\tau \norm{Ce^{As}\delta z_0}_2^2\,ds
=
\delta z_0^\top W_{o,\tau}(A,C)\delta z_0.
\]
Observability of $(C,A)$ implies
$W_{o,\tau}(A,C)\succ0$, and hence $\delta z_0=0$. Thus $\Psi(x):=\Phi(\eta)$, where $\eta$ is any admissible window reaching $x$. The coverage assumption guarantees that such a window exists, and the preceding argument guarantees that all choices give the same value. This also proves uniqueness for the fixed encoder. 

To establish smoothness, fix $\bar x\in\mathcal O$. Local existence and continuous dependence allow us to choose a neighborhood $V\subseteq\mathcal O$ of $\bar x$
and a time $\tau>0$ such that the zero-input flow $\varphi_s(x_0)$
is defined in $\mathcal D$ for all $x_0\in V$ and
$s\in[0,\tau]$. For each \(x_0\in V\), choose any admissible window reaching \(x_0\) and continue with zero input. The output equation gives
\[
h(\varphi_s(x_0))=Ce^{As}\Psi(x_0).
\]
Multiplication by $e^{A^\top s}C^\top$ and integration yield
\begin{equation}\label{eq:smooth_recovery}
\Psi(x_0)
=
W_{o,\tau}(A,C)^{-1}
\int_0^\tau e^{A^\top s}C^\top
h(\varphi_s(x_0))\,ds.
\end{equation}
Since $f$ and $h$ are $C^r$, smooth dependence of the flow
on its initial condition implies that the right-hand side
is $C^r$ in $x_0$. Therefore $\Psi\in C^r(\mathcal O;\mathbb R^N)$.

For any $x\in\mathcal O$ and constant input $v$ sufficiently
close to zero, consider a local trajectory starting from $x$.
Differentiating $z^{\mathrm{e}}=\Psi(x)$ along this trajectory and using
the exact lifted dynamics gives
\[
J_\Psi(x)\bigl(f(x)+g(x)v\bigr)=A\Psi(x)+Bv.
\]
Setting $v=0$ yields $J_\Psi(x)f(x)=A\Psi(x)$.
Subtracting this identity and varying $v$ in a neighbourhood
of zero gives $J_\Psi(x)g(x)=B$.
Finally, the output equation gives $h(x)=C\Psi(x)$.
\end{proof}


The conclusion of Proposition \ref{prop:encoder_factorization} is that input-output windows
reaching the same nonlinear state must yield the same encoded
state when their future output responses are exact and the lifted
realization is observable.

\begin{proposition}[Approximate consistency of encoded states]
\label{prop:code_disagreement}
Consider two input-output windows $\eta_1$ and $\eta_2$
whose associated nonlinear trajectories satisfy
$x_1(0)=x_2(0)=x_0$.
Under a common admissible future input on $[0,T]$, $T>0$,
suppose that the nominal model initialized at
$\Phi_\theta(\eta_i)$ has output-prediction error at most
$\bar\varepsilon$ in $L^2(0,T)$, for $i=1,2$.
If $W_{o,T}(A,C)\succ0$, then
\[
\|\Phi_\theta(\eta_1)-\Phi_\theta(\eta_2)\|_2
\leq
\frac{2\bar\varepsilon}
{\sqrt{\lambda_{\min}(W_{o,T})}}.
\]
\end{proposition}

\begin{proof}
Since the nonlinear trajectories start from the same state
$x_0$ and receive the same future input, uniqueness of
solutions gives a common future output $y$.
For $i=1,2$, the nominal predictions and their errors are
\[
\dot{\hat z}_i=A\hat z_i+Bu,
\;
\hat z_i(0)=\Phi_\theta(\eta_i),
\;
\hat y_i=C\hat z_i,
\;
\varepsilon_i=y-\hat y_i,
\]
where the assumed error bounds mean
\[
\|\varepsilon_i\|_{L^2(0,T)}
:=
\left(
\int_0^T\|\varepsilon_i(t)\|_2^2\,dt
\right)^{1/2}
\leq\bar\varepsilon.
\]

Let $\delta z_0:=\Phi_\theta(\eta_1)-\Phi_\theta(\eta_2)$.
The common input cancels when subtracting the nominal
dynamics, so
\[
Ce^{At}\delta z_0
=
\hat y_1(t)-\hat y_2(t)
=
\varepsilon_2(t)-\varepsilon_1(t).
\]
Consequently, using the definition of the observability  Gramian
\begin{equation*}
\begin{split}
\lambda_{\min}(W_{o,T})\|\delta z_0\|_2^2
&\leq \delta z_0^\top W_{o,T}\delta z_0
\\&=\int_0^T\|Ce^{At}\delta z_0\|_2^2\,dt
\\&=\|\varepsilon_2-\varepsilon_1\|_{L^2(0,T)}^2.
\end{split}
\end{equation*}
Taking square roots and applying the triangle inequality
gives
\begin{equation}
\begin{split}
\|\delta z_0\|_2
&\leq
\frac{
\|\varepsilon_1\|_{L^2(0,T)}
+
\|\varepsilon_2\|_{L^2(0,T)}
}{
\sqrt{\lambda_{\min}(W_{o,T})}
}
\\&\leq
\frac{2\bar\varepsilon}
{\sqrt{\lambda_{\min}(W_{o,T})}}.
\label{eq:code_disagreement_bound}
\end{split}
\end{equation}
Thus, the encoded states differ in Euclidean norm by at most $2\bar\varepsilon/
\sqrt{\lambda_{\min}(W_{o,T})}$.
In particular, if both output-prediction errors vanish then the encoded states are equal:
$\Phi_\theta(\eta_1)=\Phi_\theta(\eta_2)$.
\end{proof}

Proposition~\ref{prop:code_disagreement} bounds the difference between encodings of windows associated with the same current nonlinear state. The bound becomes smaller when prediction
errors decrease or the smallest observability Gramian eigenvalue increases. 

\subsection{Input-map compatibility of the lifted realization}
\label{subsec:additive_compatibility}
\label{sec:input_structure}

We next examine whether the assumption of a constant input map $B$ in the
lifted realization~\eqref{eq:problem_target_model_ct} is compatible with the
nonlinear control action of the original system~\eqref{eq:nonlinear_ct}.
Let $z=\Psi(x)$ be a differentiable state-based lifting and define its
Jacobian by
\begin{equation}
    J_\Psi(x)
    :=
    \frac{\partial \Psi}{\partial x}(x)
    \in\R^{N\times n}.
\end{equation}
Since $z=\Psi(x)$, the chain rule gives $\dot z
    =
    J_\Psi(x)\dot x$. Substituting the original control-affine dynamics
\eqref{eq:nonlinear_ct} yields
\begin{equation*}
    \dot z
    =
    \underbrace{J_\Psi(x)f(x)}_{
        \substack{\text{actual autonomous dynamics}\\
        \text{in the lifted space}}}
    +
    \underbrace{J_\Psi(x)g(x)}_{
        \substack{\text{actual input map}\\
        \text{in the lifted space}}}
    u.
\end{equation*}
On the other hand, the additive lifted realization
\eqref{eq:exact_additive_lti_for_obs} assumes
\[
    \dot z
    =
    \underbrace{A\Psi(x)}_{
        \substack{\text{assumed linear dynamics}\\
        \text{in the lifted space}}}
    +
    \underbrace{B}_{
        \substack{\text{assumed constant input}\\
        \text{map in the lifted space}}}
    u.
\]

Let $\mathcal O\subseteq\mathcal X$ be an open set and $z=\Psi(x)$ a continuously differentiable state-based lifting on $\mathcal O$. For the additive lifted realization to reproduce the nonlinear dynamics exactly on $\mathcal O$, these two expressions for
$\dot z$ must agree for every $x\in\mathcal O$ and every admissible input. If the admissible input set contains a neighbourhood of zero, their autonomous and input-dependent terms must therefore match separately:
\begin{subequations}\label{eq:exact_conditions}
\begin{align}
    J_\Psi(x)f(x)
    &=
    A\Psi(x),
    \label{eq:autonomous_closure}\\
    J_\Psi(x)g(x)
    &=
    B.
    \label{eq:input_straightening}
\end{align}
\end{subequations}
The first relation requires the actual autonomous dynamics in the lifted
space to match the assumed linear dynamics $A\Psi(x)$. The second
requires the actual input map in the lifted space, $J_\Psi(x)g(x)$, to be
independent of the state and therefore representable by the constant matrix
$B$. Writing $B=[\,b_1\ \cdots\ b_m\,]$, the exact matrix relation \eqref{eq:input_straightening} becomes
\begin{equation}
    J_\Psi(x)g_j(x)=b_j,
    \qquad
    j=1,\ldots,m.
    \label{eq:input_straightening_columns}
\end{equation}
The compatibility condition
\eqref{eq:input_straightening_columns} also has a direct interpretation in terms of the input vector field. For a single-input system, let $\phi_g^s(x)$ denote the solution at parameter $s$ of
\[
\frac{d\xi}{ds}=g(\xi(s)),
\qquad \xi(0)=x.
\]

\begin{lemma}[Input-map compatibility along the input flow]
The condition
\[
J_\Psi(x)g(x)=b
\]
holds on $\mathcal O$ if and only if
\begin{equation}
\Psi(\phi_g^s(x))=\Psi(x)+sb
\label{eq:input_flow_compatibility}
\end{equation}
for every $x\in\mathcal O$ and every $s$ for which the
trajectory segment between parameters $0$ and $s$ remains
in $\mathcal O$.
\end{lemma}

\begin{proof}
By the chain rule,
\[
\frac{d}{ds}\Psi(\phi_g^s(x))
=
J_\Psi(\phi_g^s(x))g(\phi_g^s(x)).
\]
If $J_\Psi g=b$, integrating from $0$ to $s$ gives
\[
\Psi(\phi_g^s(x))-\Psi(x)
=
\int_0^s b\,d\sigma
=
sb
\]
which is exactly equation \eqref{eq:input_flow_compatibility}. Conversely, differentiating
\eqref{eq:input_flow_compatibility} at $s=0$ gives
$J_\Psi(x)g(x)=b$.
\end{proof}

Thus, along every trajectory generated by $g$, the lifted
state changes at the constant rate $b$. Its image is a
straight line, or a single point when $b=0$.
The lifting can remain nonlinear as a function of $x$,
provided that this constant-rate condition holds along
the input flow. For multiple inputs, the condition applies
separately to each input vector field $g_j$ and the
corresponding column $b_j$ of $B$.

\begin{definition}[Residuals of the additive lifted realization]
For a continuously differentiable state-based lifting $\Psi$
and constant matrices $A$, $B$, and $C$, define
\begin{subequations}
\begin{align}
e_A(x)
&:=J_\Psi(x)f(x)-A\Psi(x),\\
E_B(x)
&:=J_\Psi(x)g(x)-B,\\
e_y(x)
&:=h(x)-C\Psi(x).
\end{align}
\end{subequations}
Here, $e_A(x)\in\mathbb R^N$,
$E_B(x)\in\mathbb R^{N\times m}$, and
$e_y(x)\in\mathbb R^p$ are the autonomous-dynamics,
input-map, and output residuals, respectively.
\end{definition}

With $z=\Psi(x)$, the nonlinear dynamics and output can
therefore be written as
\begin{align*}
&\dot z=Az+Bu+d(x,u),\\
&y=Cz+e_y(x).    
\end{align*}
where $d(x,u):=e_A(x)+E_B(x)u$. The residual $E_B(x)$ measures the difference between the actual lifted input map $J_\Psi(x)g(x)$ and the assumed constant matrix $B$. Its contribution to the lifted dynamics
is $E_B(x)u$, which vanishes when $u=0$. Consequently, autonomous trajectories alone cannot reveal this input-map mismatch through the dynamics.

Let $\hat z$ denote the nominal trajectory generated by the assumed
additive lifted realization \eqref{eq:exact_additive_lti_for_obs} from the same initial lifted state and under the same applied input. Variation of constants then gives $z(t)-\hat z(t)
    =
    \int_0^t
    e^{A(t-s)}
    \left(
        e_A(x(s))+E_B(x(s))u(s)
    \right)\,ds$. Using the Euclidean vector norm $\|\cdot\|_2$ and the corresponding induced
matrix $2$-norm,
\begin{equation} \label{eq:residual_error_bound}
\begin{split}
& \|z(t)-\hat z(t)\|_2
\le
\int_0^t
\|e^{A(t-s)}\|_2
\Bigl(
    \|e_A(x(s))\|_2
    +\\
&\qquad\qquad\qquad\qquad\quad
\|E_B(x(s))\|_2\,\|u(s)\|_2
\Bigr)\,ds .   
\end{split}
\end{equation}
Accordingly, $\|E_B(x)\|_2$ provides a state-dependent measure of the discrepancy between the actual input map in the lifted space and the assumed constant input map. Equation~\eqref{eq:residual_error_bound} shows how autonomous-dynamics and input-map residuals accumulate in the lifted-state prediction. The contribution of the input-map residual also depends on the applied input through $E_B(x)u$. Direct evaluation of $E_B$ requires the nonlinear input map $g$ and a differentiable state-based lifting $\Psi$. When these maps are known, $J_\Psi(x)$ can be computed analytically or by automatic differentiation and multiplied by $g(x)$. Equivalently, the $j$th column is the directional derivative of $\Psi$ along $g_j$:
\[
J_\Psi(x)g_j(x)
=\left.\frac{d}{ds}\Psi\bigl(\phi_{g_j}^{s}(x)\bigr)
\right|_{s=0},
\]
where $\phi_{g_j}^{s}$ denotes the local flow of $g_j$. The analytical lift in Section~\ref{subsec:S1} illustrates this
calculation. For a general window encoder, differentiation with respect to its window entries gives $D_\eta\Phi_\theta$, rather than $J_\Psi$, and therefore does not by itself provide
$J_\Psi g$. Direct evaluation of $E_B$ is not required by the input-output identification procedure; it describes the structural mismatch when a differentiable state-based lifting and the nonlinear input map are available.

A nonzero $E_B$ indicates a mismatch with the chosen constant input matrix, but does not determine a unique alternative model structure. If $\Psi$ is injective on $\mathcal O$,
the actual input map can be expressed on $\Psi(\mathcal O)$ as a function of the lifted state:
\[
\mathcal B(\Psi(x)):=J_\Psi(x)g(x).
\]
The lifted dynamics then satisfy
\[
\dot z=Az+\mathcal B(z)u+e_A(x).
\]
Under exact autonomous closure, $e_A=0$. A bilinear representation additionally requires
\[
\mathcal B(z)
=
B+
\begin{bmatrix}
N_1z & \cdots & N_mz
\end{bmatrix},
\]
which gives
\[
\dot z=Az+Bu+\sum_{j=1}^m N_jz\,u_j.
\]
Bilinear Koopman models have been studied by \citet{goswami2017bilinear,peitz2020generators,otto2024bilinear}. In addition, \citet{strasser2026io} construct a bilinear surrogate using an extended state formed from input-output trajectories.
The present framework retains the additive model and uses $E_B$ to quantify its input-map mismatch.

\paragraph{Effect on output separation.}
For two state-based trajectories under a common input, let $\delta d$ and $\delta e_y$ denote residual differences. Their true output difference is $\delta y(t)=Ce^{At}\delta z_0+\epsilon(t)$ with
 $\epsilon(t)=\int_0^t Ce^{A(t-s)}\delta d(s)\,ds+\delta e_y(t)$. The reverse triangle inequality yields
\begin{equation}\label{eq:separation_with_error}
 \norm{\delta y}_{L^2}\ge
 \sqrt{\delta z_0^\top\WoT\delta z_0}-\norm{\epsilon}_{L^2}.
\end{equation}
If additionally $\norm{\epsilon}_{L^2}\le\kappa_T\norm{\delta z_0}$ on a specified domain, the separation margin is at least $\sqrt{\lambda_{\min}(\WoT)}-\kappa_T$ in lifted coordinates. Such a proportional bound is an additional assumption, not a consequence of the
empirical learning constraint below. It explains why nominal observability must be interpreted together with predictive fidelity over the same horizon. These residuals also affect output prediction. Accordingly, observability of the nominal pair $(C,A)$ must be assessed together with the accuracy of the lifted dynamics and output map.

\subsection{Controllability of the lifted realization}
\label{subsec:controllability}

Having assessed when the constant matrix $B$ provides a compatible
representation of the nonlinear input action in the lifted space, we now consider the controllability of the resulting finite-dimensional realization. For \eqref{eq:exact_additive_lti_for_obs}, define the finite-horizon controllability Gramian
\begin{equation}
    W_{\mathrm{c},T}(A,B)
    :=
    \int_0^T
    e^{At}BB^{\top}e^{A^{\top}t}\,dt
    \label{eq:wc_ct}
\end{equation}
and the Kalman controllability matrix $\mathcal C(A,B)
    :=
    \begin{bmatrix}
        B&AB&\cdots&A^{N-1}B
    \end{bmatrix}$. The Gramian \eqref{eq:wc_ct} quantifies how strongly different lifted-state directions can
be influenced by the input over the finite horizon $[0,T]$. Directions
associated with small eigenvalues of $W_{\mathrm{c},T}$ require comparatively large
input energy to excite, while a zero eigenvalue indicates a direction that
cannot be reached through the input over that horizon. The relation to the nonlinear control-affine system can be made explicit under exact closure. For vector fields $f$ and $g$, define the Lie bracket by
\begin{equation}
    [f,g]:=J_gf-J_fg,
    \label{eq:lie_bracket}
\end{equation}
the same definition being used for any pair of vector fields, and recursively define
$\operatorname{ad}_f^0g=g$ and
$\operatorname{ad}_f^{q+1}g=[f,\operatorname{ad}_f^qg]$.

\begin{proposition}[Relation between nonlinear Lie brackets and lifted controllability]
\label{prop:bracket_krylov}
Assume that $\Psi$, $f$, and $g_j$ are sufficiently smooth on an open set
$\mathcal{O}\subseteq\mathcal{X}$ and that exact autonomous closure and exact
input-map compatibility hold on $\mathcal{O}$, namely
\[
    J_\Psi(x)f(x)=A\Psi(x),
    \;
    J_\Psi(x)g(x)=B,
    \; x\in\mathcal{O}.
\]
Then, for each input vector field $g_j$ and every nonnegative integer $q$ for which the iterated Lie bracket is defined on $\mathcal{O}$,
\begin{equation}
    J_\Psi(x)\operatorname{ad}_f^q g_j(x)
    =
    (-1)^q A^q b_j,
    \;
    x\in\mathcal{O},
    \label{eq:bracket_krylov}
\end{equation}
where $b_j$ denotes the $j$th column of $B$.
\end{proposition}

\begin{proof}
For $q=0$, the identity follows directly from
input-map compatibility:
\[
J_\Psi(x)g_j(x)=b_j.
\]

Suppose that the identity holds for some $q\geq0$ and set $X:=\operatorname{ad}_f^q g_j$. The induction hypothesis gives
\[
L_X\Psi
=
J_\Psi X
=
(-1)^q A^q b_j.
\]
Since $A$ and $b_j$ are constant, this expression is independent of $x$, and therefore $L_f(L_X\Psi)=0$. Exact autonomous closure also gives $L_f\Psi=A\Psi$. Using
$\operatorname{ad}_f^{q+1}g_j=[f,X]$
and the Lie-derivative identity associated with the
chosen bracket convention, we obtain
\begin{align*}
&J_\Psi\operatorname{ad}_f^{q+1}g_j
=J_\Psi[f,X]
=L_f(L_X\Psi)-L_X(L_f\Psi)=\\
&=-A L_X\Psi
=-A\bigl((-1)^q A^q b_j\bigr)
=(-1)^{q+1}A^{q+1}b_j.
\end{align*}
The result follows by induction.
\end{proof}

The preceding proposition concerns repeated brackets with $f$ and does not, by itself, characterize nonlinear controllability. It does, however, impose a rank restriction on the lifted realization. This restriction is particularly relevant for an overcomplete lifting, by which we mean $N>n$.

\begin{corollary}[Controllability limitation of an overcomplete lift]
\label{cor:controllability_limitation}
Assume that $\Psi,f,g_1,\ldots,g_m$ are of class $C^N$ on
the open set $\mathcal O$ and that the exact closure and
input-map compatibility conditions~\eqref{eq:exact_conditions}
hold there. Then the identities of
Proposition~\ref{prop:bracket_krylov} apply for
$q=0,\ldots,N-1$, and the lifted controllability matrix satisfies
\[
\operatorname{range}\mathcal C(A,B)
\subseteq
\operatorname{range}J_\Psi(x),
\qquad x\in\mathcal O.
\]
Consequently,
\begin{equation}
\operatorname{rank}\mathcal C(A,B)
\leq
\operatorname{rank}J_\Psi(x)
\leq n.
\label{eq:lifted_controllability_rank}
\end{equation}
In particular, if $N>n$, the exact additive lifted realization
is not fully controllable in $\mathbb R^N$.
\end{corollary}

\begin{proof}
For every $j=1,\ldots,m$ and $q=0,\ldots,N-1$, the preceding proposition gives
\[
A^q b_j
=
(-1)^q J_\Psi(x)\operatorname{ad}_f^q g_j(x)
\in
\operatorname{range}J_\Psi(x).
\]
These vectors are the columns of $\mathcal C(A,B)$. Their span is therefore contained in
$\operatorname{range}J_\Psi(x)$, which has dimension at most $n$.
\end{proof}

Under the assumptions of Corollary \ref{cor:controllability_limitation}, an exact additive lift with $N>n$ cannot be fully controllable. Consequently, $W_{c,T}$ is singular and $\log\det W_{c,T}$ cannot provide a finite learning objective. We therefore assess controllability as a structural property and promote observability subject to prediction and consistency constraints. The rank bound need not hold for approximate models or general window-based predictors. For nominal stabilization, the relevant requirement is \textit{stabilisability}: all uncontrollable eigenvalues of $A$ must have negative real parts. This alone does not guarantee stability of the nonlinear closed loop when the lifted model is approximate.

\section{Identification and output-feedback control}
\label{sec:learning}

In this section we develop a computational procedure for learning an additive
lifted model of the form \eqref{eq:exact_additive_lti_for_obs} from measured input-output trajectories and then using it for output-feedback control. We initially construct lifted coordinates from
input-output windows and normalize their scale (Section~\ref{sec:output_history}). For a fixed encoder, we then identify
the model matrices by integral regression (Section~\ref{sec:integral}). The encoder is initially trained for output prediction and consistency with the identified dynamics, and subsequently refined to promote finite-horizon observability while keeping these two errors within separate prescribed bounds (Section~\ref{sec:obs}). Each encoder update entails recomputing the normalization and identifying the corresponding model matrices. Finally, the learned model is used for online lifted-state estimation and predictive control
(Section~\ref{sec:ofc}). 


\subsection{Lifted-state construction and normalization}
\label{sec:output_history}

For the causal input-output window~\eqref{eq:history_vector_ct}, define
$z^{\mathrm{e}}(t)=\Phi_\theta(\eta(t))\in\R^N$, where the superscript $e$ denotes
the encoded state, and seek a representation satisfying
\begin{equation}\label{eq:lifted_model}
 \dot z^{\mathrm{e}}\approx Az^{\mathrm{e}}+Bu,\qquad \hat y=Cz^{\mathrm{e}}.
\end{equation}
Let $F_\theta$ be the neural-network map specified in~\ref{app:implementation}, with parameters $\theta$, and set
$\Phi_\theta(\eta)=F_\theta(\eta)-F_\theta(0)$. This enforces
$\Phi_\theta(0)=0$: the nominal zero input-output window at the chosen
equilibrium maps to the lifted origin, which is an equilibrium of
\eqref{eq:lifted_model} for $u=0$. It does not impose stability.
The interpretation of $\Phi_\theta$ as a state-based lifting, i.e., the factorization~\eqref{eq:factorization}, requires the
conditions of Proposition~\ref{prop:encoder_factorization};
Proposition~\ref{prop:code_disagreement} addresses approximate consistency.

Comparing observability across encoders also requires fixing the scale of the
lifted coordinates, since the Gramian~\eqref{eq:ct_obs_gramian} depends on the choice of coordinates \citep{moore1981principal}.
For example, the scaled coordinates $z' = az^{\mathrm{e}}$, $a>0$, with $B' = aB$ and
$C' = a^{-1}C$, give identical output predictions but
$W_{o,T}(A,C') = a^{-2}W_{o,T}(A,C)$. An encoder could thus increase
$\log\det W_{o,T}$ by $-2N\log a$ merely by shrinking its output, without
any gain in distinguishability. We prevent this by normalizing the encoded
states by their covariance $\Sigma_z$ on the identification data \citep{kessy2018optimal}. Choose
$\beta\ge 0$ with $\Sigma_z+\beta I_N\succ 0$, let $L_\beta$ be the Cholesky
factor of $\Sigma_z+\beta I_N$, and set $z_n = L_\beta^{-1}z^{\mathrm{e}}$ such that
\begin{equation}
\label{eq:normalized}
 A_n = L_\beta^{-1}AL_\beta,\;
  B_n = L_\beta^{-1}B,\; C_n = CL_\beta.
\end{equation}
Substituting $z^{\mathrm{e}} = L_\beta z_n$ into~\eqref{eq:lifted_model} shows that this
change of coordinates preserves the model's input-output behaviour. The sample mean is used to compute $\Sigma_z$ but is not subtracted from $z^{\mathrm{e}}$,
so the equilibrium remains at the origin. If $\Sigma_z\succ 0$ and $\beta=0$,
the normalized coordinates have identity covariance, so encoders related by an
invertible linear change of coordinates differ after normalization only by an
orthogonal transformation and share the same Gramian eigenvalues; the criterion
then compares representations rather than coordinate choices. A positive
$\beta$ gives up this exact invariance but limits the amplification of
directions in which the encoded states barely vary, since
$\|L_\beta^{-1}\|_2\le\beta^{-1/2}$. We fix $N$ within each model comparison
and examine the effect of normalization in Section~\ref{sec:duffing_results}.

\subsection{Integral identification}
\label{sec:integral}

For fixed $\theta$, the normalized samples are known and the matrices
$A_n,B_n,C_n$ are fitted by regression. Integrating
$\dot z_n\approx A_nz_n+B_nu$ over $[t_i,t_{i+1}]$ gives
\begin{equation}\label{eq:integral}
\begin{aligned}
 &z_{n,i+1}-z_{n,i}\approx A_n\xi_i+B_n\nu_i,\\
 &\xi_i=\int_{t_i}^{t_{i+1}}z_n(t)\,dt,\;
 \nu_i=\int_{t_i}^{t_{i+1}}u(t)\,dt.
\end{aligned}
\end{equation}
For sampling interval $\Delta t$, the trapezoidal approximation is
$\xi_i\approx\frac{\Delta t}{2}(z_{n,i}+z_{n,i+1})$.
With zero-order-held inputs, meaning $u(t)=u_i$ on each sampling interval,
$\nu_i=\Delta t\,u_i$ exactly. This formulation avoids constructing
numerical derivative targets \citep{garnier2008identification}.

Collect the increments and integrals as columns of
$\Delta Z_n$, $Z_I$, and $U_I$, respectively, and set
$\Omega_n=\col(Z_I,U_I)\in\R^{(N+m)\times M}$, where $M$ is the
number of fitting intervals. Then
$\Delta Z_n\approx[A_n\ B_n]\Omega_n$.
Likewise, collect the $M_y$ encoded-state and measured-output samples,
including final endpoints, in $Z_n$ and $Y$, so that $Y\approx C_nZ_n$.
The regularized least-squares fits are
\begin{align}\label{eq:AB_ls}
& \min_{A_n,B_n} 
 \frac1M\norm{\Delta Z_n-[A_n\ B_n]\Omega_n}_F^2
 +\lambda_{AB}\norm{[A_n\ B_n]}_F^2,\\
 &\min_{C_n} 
 \frac1{M_y}\norm{Y-C_nZ_n}_F^2+\lambda_C\norm{C_n}_F^2.   
\end{align}
The first terms measure average squared fitting errors; the penalties
limit the size of the fitted coefficients. For positive regularization
parameters, the normal equations give
\begin{align*}
&[A_n\ B_n]
 =\Delta Z_n\Omega_n^\top
   (\Omega_n\Omega_n^\top+M\lambda_{AB}I_{N+m})^{-1},\\
&C_n=YZ_n^\top(Z_nZ_n^\top+M_y\lambda_C I_N)^{-1}.   
\end{align*}   
All fits use the identification data, and $C_n$ maps to physical output
units. A unique unregularized estimate of $A_n,B_n$ requires
$\rank\Omega_n=N+m$. We assess relative conditioning using
\begin{equation}\label{eq:informativity}
 \chi_I=\frac{\sigma_{\min}(\Omega_n)}{\sigma_{\max}(\Omega_n)}.
\end{equation}
For full row rank, this is the reciprocal of the spectral condition
number: a small value indicates weakly represented regressor directions
and sensitivity to data perturbations. Regularization can select a unique
fit when rank is missing, but cannot supply the missing information. For example, under continuous fixed feedback $u(t)=Kz_n(t)$,
integration gives $\nu_i=K\xi_i$ and hence $U_I=KZ_I$. Therefore,
\begin{equation}\label{eq:rank_fb}
 \Omega_n=\begin{bmatrix}I_N\\K\end{bmatrix}Z_I,
 \qquad \rank\Omega_n\le N.
\end{equation}
The dynamics reduce to $\dot z_n\approx(A_n+B_nK)z_n$, so the data
cannot distinguish $A_n$ and $B_n$ without further restrictions
\citep{vandenhof1995identification,forssell1999closedloop}.
An additional probing input, i.e., in $u=Kz_n+v_{\rm exc}$,
can provide independent input variation; the resulting regression matrix
must still have full row rank. Factorization~\eqref{eq:rank_fb}
need not hold for sample-and-hold feedback, whose actual regressors must
be assessed. The data must also cover the intended operating region.
These are requirements on the identification experiment, distinct from
controllability of the realization \citep{vanwaarde2020data}.
Measurement errors perturb both the regressors and the increments, so
integral regression may still produce biased estimates
\citep{soderstrom2007errors}.

\subsection{Observability optimization under accuracy constraints}
\label{sec:obs}

To assess predictions over $T_p=H\Delta t$, we use the exact sampled
form of the nominal linear model under zero-order-held inputs,
$z_{k+1}=A_dz_k+B_du_k$, where
\begin{equation}\label{eq:zoh}
 A_d=e^{A_n\Delta t},\qquad
 B_d=\int_0^{\Delta t}e^{A_ns}B_n\,ds.
\end{equation}
For each starting index $s$ in a set $\mathcal S$, initialize from the
measurement window and propagate using the recorded input sequence:
\begin{equation}\label{eq:training_rollout}
 \hat z_{s|s}=z_{n,s},\qquad
 \hat z_{s+j+1|s}=A_d\hat z_{s+j|s}+B_du_{s+j}.
\end{equation}
Here $\hat z_{s+j|s}$ is the prediction at time $s+j$ initialized at
time $s$; subsequent measurements do not correct this prediction.
Define the output-prediction and lifted-state consistency errors by
\begin{align}
 \Lpred&=\frac{1}{|\mathcal S|(H+1)p}
 \sum_{s\in\mathcal S}\sum_{j=0}^{H}
 \norm{D_y(y^{\rm m}_{s+j}-C_n\hat z_{s+j|s})}^2,
 \label{eq:lroll}\\
 \Lcons&=\frac{1}{|\mathcal S|HN}
 \sum_{s\in\mathcal S}\sum_{j=1}^{H}
 \norm{z_{n,s+j}-\hat z_{s+j|s}}^2.
 \label{eq:consistency_loss}
\end{align}
The output scaling is fixed from the identification data as
\begin{equation}\label{eq:Dy}
 D_y=\operatorname{diag}\bigl(s_1^{-1},\ldots,s_p^{-1}\bigr),
\end{equation}
where $s_i$ is the sample standard deviation of the measured output channel
$y^{\rm m}_i$ over all identification samples, so that each channel
contributes errors in units of its own variation; for the scalar outputs of
Section~\ref{sec:examples}, $D_y=1/s_1$. The same $D_y$ scales the output
entries of the encoder input (for details see~\ref{app:implementation}) and is held
fixed during both training stages. The denominators average over starting windows, prediction
samples, and coordinates. The first loss measures output agreement,
including the initial readout at $j=0$. The second compares model
propagation with the coordinates encoded from subsequent windows; its
initial term vanishes by construction. Thus it assesses whether the
encoder and the identified dynamics give consistent lifted trajectories.
Neither loss requires measured nonlinear states. Both are needed: output
agreement alone leaves poorly observed lifted directions weakly constrained,
whereas consistency alone is achieved by uninformative representations, such
as an encoder that always returns zero. First minimize $\Lpred+\lambda_c\Lcons$, with $\lambda_c>0$, to
obtain a reference model. We then improve observability within prescribed
error limits relative to this reference. With $\widetilde C_n=D_yC_n$,
define the sampled finite-horizon Gramian
\begin{equation}\label{eq:sampled_gramian}
 W_{o,H}^{\Delta}=\Delta t\sum_{j=0}^{H-1}
 (A_d^j)^\top\widetilde C_n^\top\widetilde C_nA_d^j.
\end{equation}
By Lemma~\ref{lem:known_input_cancellation}, an initial perturbation
$\delta z_0$ under a common known input produces the scaled output
perturbation $\widetilde C_nA_d^j\delta z_0$ at sample $j$.
Its accumulated squared magnitude over these $H$ samples is
$\Delta t\sum_{j=0}^{H-1}\norm{\widetilde C_nA_d^j\delta z_0}^2
=\delta z_0^\top W_{o,H}^{\Delta}\delta z_0$.
Equation~\eqref{eq:sampled_gramian} also gives a left-endpoint quadrature
approximation to the continuous-time Gramian with output map
$\widetilde C_n$ over $[0,T_p]$. Starting from the reference model, solve
\begin{equation}\label{eq:obsloss}
\begin{aligned}
 \min_\theta\quad &
 \ell_{\rm obs}=-\log\det(W_{o,H}^{\Delta}+\varepsilon I_N)\\
 \text{s.t.}\quad
 &\Lpred\le b_p:=(1+\delta_p)\Lpred^{\rm ref}+\tau_p,\\
 &\Lcons\le b_c:=(1+\delta_c)\Lcons^{\rm ref}+\tau_c.
\end{aligned}
\end{equation}
The objective equals $-\sum_{i=1}^N\log(\lambda_i+\varepsilon)$,
where $\lambda_i$ are the Gramian eigenvalues and $\varepsilon>0$
keeps the determinant positive. The parameters $\delta_p,\delta_c\ge0$
specify allowed relative increases in the reference losses, while
$\tau_p,\tau_c\ge0$ provide absolute tolerances, including when a
reference loss is close to zero. Separate bounds prevent an improvement
in one error from compensating for a violation of the other bound.
These empirical bounds apply to the identification data; they do not
establish trajectory-wise error bounds or guarantee accuracy on new data. For each $\theta$, the encoded samples, normalization, and regression
fits are recomputed. Thus $A_n,B_n,C_n$ depend on $\theta$, and this
dependence is included when differentiating the objective and constraints
\citep{golub2003separable}. We retain the feasible iterate with the smallest
$\ell_{\rm obs}$; if none improves on the reference, the reference is
returned. Independent validation data select the reference model and
design parameters. The constrained stage imposes no additional validation
bound and carries no global-optimality guarantee. The linear part of $\Phi_\theta$ is initialized by subspace identification
from the same input-output records \citep{vanoverschee1996subspace}.
Future inputs and outputs are used only in this offline initialization;
evaluating $\Phi_\theta$ requires only the past and present window.
Random initialization is examined separately in Section~\ref{subsec:S1}.
Implementation settings are given in~\ref{app:implementation}.
The following result interprets the objective through recovery of an
initial lifted state from a finite set of nominal observations.

\begin{proposition}[Observability measure and lifted-state recovery]
\label{prop:posterior}
Consider the normalized sampled realization over $H$ samples, after
subtracting the known-input response as in
Lemma~\ref{lem:known_input_cancellation}:
\[
 \widetilde y_j=\widetilde C_nA_d^jz_0+\widetilde v_j,
 \; j=0,\ldots,H-1.
\]
Assume independent noise $\widetilde v_j\sim\mathcal N(0,\sigma_s^2I_p)$
and an independent initial state $z_0\sim\mathcal N(0,I_N)$.
For $\varepsilon=\sigma_s^2\Delta t>0$, the posterior covariance is
\[
 P=\varepsilon(W_{o,H}^{\Delta}+\varepsilon I_N)^{-1},
 \;
 \log\det P=\ell_{\rm obs}+N\log\varepsilon.
\]
Without the prior, if $W_{o,H}^{\Delta}\succ0$, the least-squares estimate
of $z_0$ has error covariance $\varepsilon(W_{o,H}^{\Delta})^{-1}$.
\end{proposition}

\begin{proof}
Let $\Theta=\col(\widetilde C_n,\widetilde C_nA_d,\ldots,
\widetilde C_nA_d^{H-1})$. Stacking the measurements gives
$\widetilde Y=\Theta z_0+\widetilde V$, with
$\widetilde V\sim\mathcal N(0,\sigma_s^2I_{Hp})$ and
$\Theta^\top\Theta=W_{o,H}^{\Delta}/\Delta t$.

The negative log-posterior density, up to terms independent of $z_0$, is
$\frac12\norm{z_0}^2
 +\frac{1}{2\sigma_s^2}\norm{\widetilde Y-\Theta z_0}^2$.
Collecting its quadratic terms shows that the inverse posterior
covariance is
$P^{-1}=I_N+\frac{\Theta^\top\Theta}{\sigma_s^2}
       =I_N+\frac{W_{o,H}^{\Delta}}{\varepsilon}.$
Inversion gives the stated $P$, and
$\det P=\varepsilon^N/\det(W_{o,H}^{\Delta}+\varepsilon I_N)$
gives the log-determinant identity.

Without the prior, the least-squares error is
$(\Theta^\top\Theta)^{-1}\Theta^\top\widetilde V$, whose covariance is
$\sigma_s^2(\Theta^\top\Theta)^{-1}
=\varepsilon(W_{o,H}^{\Delta})^{-1}$.
\end{proof}

For fixed $N$ and $\varepsilon$, minimizing $\ell_{\rm obs}$ therefore
minimizes the volume of a Gaussian uncertainty ellipsoid at a fixed
probability level for this nominal initial-state estimation problem.
Along the eigenvector of $W_{o,H}^{\Delta}$ with eigenvalue $\lambda_i$, the
posterior variance is $p_i=\varepsilon/(\lambda_i+\varepsilon)$: the
measurements do not reduce the prior uncertainty in a direction with
$\lambda_i=0$, and reduce it strongly when $\lambda_i\gg\varepsilon$.
The Gaussian prior and independent
noise are assumptions of the proposition, not consequences of coordinate
normalization. Regularization keeps the criterion finite even when the
Gramian is singular, and increasing its determinant need not increase
its smallest eigenvalue: changing the eigenvalues from $(1,1)$ to $(0.5,4)$
doubles the determinant while halving $\lambda_{\min}$. The criterion consequently differs from the
bound in Proposition~\ref{prop:code_disagreement} and does not directly
describe the recursive estimator below, which also faces continuing
model errors.

\subsection{State estimation and predictive control}
\label{sec:ofc}

Initialize the lifted-state estimate once from the measurement window,
$\hat z_{k_0|k_0}=z_{n,k_0}$. At each subsequent sample, predict with
the model and then correct using the new output measurement:
\begin{equation}
\begin{aligned}\label{eq:observer}
 &\hat z_{k|k-1}=A_d\hat z_{k-1|k-1}+B_du_{k-1},\\
 &\hat z_{k|k}=\hat z_{k|k-1}
 +L_\infty(y_k^{\rm m}-C_n\hat z_{k|k-1}).   
\end{aligned}
\end{equation}
The innovation $y_k^{\rm m}-C_n\hat z_{k|k-1}$ is the discrepancy
between the measured and predicted outputs; the gain $L_\infty$ maps it
to a lifted-state correction. The steady-state Kalman gain
\citep{anderson1979optimal} uses measurement covariance
$R_v=\sigma_v^2I_p$ and effective process covariance
$Q_w=q\widehat Q+10^{-12}I_N$. Here $\widehat Q$ is the uncentered
second moment of the one-step residuals
$r_k=z_{n,k+1}-A_dz_{n,k}-B_du_k$ on the identification data, namely
$\widehat Q=M_r^{-1}\sum_k r_kr_k^\top$ over $M_r$ residual samples.
It includes model mismatch and errors in the window-based coordinates.
The scalar $q$ is selected by validation of multistep output prediction.
We check that the eigenvalues of $(I_N-L_\infty C_n)A_d$, which governs
the homogeneous nominal estimation error, lie inside the unit circle. For comparison, we also recompute the estimate at every sample using
\begin{equation}\label{eq:window_estimate}
 \hat z_{k|k}=L_\beta^{-1}\Phi_\theta(\eta_k).
\end{equation}
This estimator reuses the current measurement window, whereas
\eqref{eq:observer} recursively propagates and corrects its previous
estimate. Both use measured inputs and outputs and initialize the same
predictive controller. For a constant reference $r$, a nominal steady target must satisfy
$z_s=A_dz_s+B_du_s$, reproduce $r$, and respect the input constraint:
\begin{equation}\label{eq:target}
 (I_N-A_d)z_s=B_du_s,\; C_nz_s=r,\; u_s\in\U.
\end{equation}
The implementation computes the minimum-norm least-squares solution of
the two equalities and clips its input to the admissible interval.
The resulting $u_c$ is used only as an input-cost centre: least squares
need not satisfy inconsistent equalities, and clipping need not preserve
a solution. Target infeasibility is reported separately from tracking
failure. Whether a violated bound proves infeasibility depends on uniqueness. For
$m=p$, the matrix $\mathcal T=\begin{bmatrix}I_N-A_d&-B_d\\C_n&0\end{bmatrix}$ 
of the two equalities is square. If $1$ is not an eigenvalue of $A_d$, its
Schur complement gives $\det\mathcal T=\det(I_N-A_d)\det G(1)$, where
$G(1):=C_n(I_N-A_d)^{-1}B_d$ is the steady-state gain of the sampled model.
Hence, whenever $G(1)$ is nonsingular, the target is unique,
\begin{equation}\label{eq:target_unique}
 u_s=G(1)^{-1}r,\qquad z_s=(I_N-A_d)^{-1}B_du_s,
\end{equation}
the minimum-norm least-squares solution coincides with it, and
$u_s\notin\U$ certifies that no admissible nominal target exists for this
reference. If instead $\mathcal T$ is singular, a violation by the
minimum-norm candidate does not exclude other admissible solutions. The
implementation records the condition number of $\mathcal T$ for every model
and reference, so that the applicable case can be verified. Writing $z_j=z_{k+j|k}$ and $u_j=u_{k+j|k}$, solve over a
control horizon of $H_c$ steps, a design quantity separate from the
learning horizon $H$,
\begin{equation}\label{eq:mpc}
\begin{aligned}
 \min_{\{u_j\}}\quad &
 \sum_{j=1}^{H_c-1}\norm{C_nz_j-r}_Q^2
 +\norm{C_nz_{H_c}-r}_{Q_f}^2 + \\ & \sum_{j=0}^{H_c-1}
 \left(\norm{u_j-u_c}_R^2+\norm{\Delta u_j}_{R_\Delta}^2\right)\\
 \text{s.t.}\quad &z_{j+1}=A_dz_j+B_du_j,\; z_0=\hat z_{k|k},\\ 
 & u_j\in\U,
 \; j=0,\ldots,H_c-1.
\end{aligned}
\end{equation}
Here $\norm{a}_Q^2=a^\top Qa$, $Q,Q_f,R_\Delta\succeq0$, and $R\succ0$.
The cost penalizes output errors along the horizon and at its endpoint,
input deviations from $u_c$, and input changes. For $j=0$, $\Delta u_j$
is measured from the previously applied input; for $j\ge1$, it equals
$u_j-u_{j-1}$. Linear prediction dynamics and convex input constraints
make this a convex quadratic program. Only the first optimized input is
applied; estimation and optimization are repeated at the next sample
\citep{korda2018linear,rawlings2017model}. The controller uses no
terminal-state constraint.

To relate prediction errors to Section~\ref{subsec:additive_compatibility},
consider a state-based lifting with
$\dot z_n=A_nz_n+B_nu+d_n$, where
$d_n=L_\beta^{-1}(e_A+E_Bu)$. A nominal prediction under the same input,
without further measurement corrections, has error dynamics
$\dot e_z=A_ne_z+d_n$, where $e_z=z_n-\hat z_n$.
Variation of constants and the output residual give
\begin{equation}
\begin{aligned}\label{eq:errdecomp}
&y^{\rm m}(t)-\hat y(t)
 =C_ne^{A_nt}\bigl(z_n(0)-\hat z_n(0)\bigr)+\\
&\int_0^t C_ne^{A_n(t-s)}d_n(s)\,ds+e_y(t)+v(t).   
\end{aligned}
\end{equation}
The four terms represent initialization, accumulated dynamics,
output-map, and measurement errors. Window-based dynamics residuals
also include inconsistency between the encoder and model propagation.
Forced predictions at varying input amplitudes assess the additive
approximation without separately identifying $e_A$ and $E_B$.
The decomposition also explains why improved nominal observability need not
yield better observer-based estimation or tracking: $W_{o,H}^{\Delta}$
describes sensitivity to the initial lifted state, whereas prediction and
control also depend on the remaining terms in~\eqref{eq:errdecomp}.
Section~\ref{sec:examples} assesses how far that improvement carries through
to estimation and control.

\begin{remark}\label{rem:mpc_guarantees}
For nonempty $\U$, the nominal input-only problem~\eqref{eq:mpc}
admits a feasible input sequence at every step. This does not establish
feasibility of the steady target~\eqref{eq:target}, nonlinear closed-loop
stability, or zero steady-state tracking error. The empirical learning
constraints provide no such guarantees; stability and tracking guarantees
require additional controller assumptions and model-error bounds
\citep{zhang2022robust,strasser2026io}.
\end{remark}
\section{Numerical examples}
\label{sec:examples}

The numerical examples examine three questions: whether identification
from input-output windows recovers a known state-based lifting, whether
observability optimization reduces dependence on past trajectories, and
whether an improved nominal Gramian is accompanied by better estimation
and output-feedback tracking. The first system admits an exact additive
lift, so its analytical realization provides a reference for the learned
coordinates. The controlled Duffing system admits no informative exact
additive lift under the stated assumptions, and therefore tests the
method in an intrinsically approximate setting.

Identification uses measured inputs and outputs. Nonlinear states are
available in simulation for data generation and evaluation, but are not
supplied as training targets to the encoder or the integral regression.
We distinguish the error constraints imposed on identification data from
performance on independent test trajectories. The reported standard
deviations describe variation across identification runs; they are not
confidence intervals or tests of statistical significance.

\subsection{System with an exact additive lift}
\label{subsec:S1}

Consider
\begin{equation}\label{eq:slow-manifold}
 \dot x_1=a_1x_1,\;
 \dot x_2=a_2(x_2-x_1^2)+u,\;
 y=x_1+x_2,
\end{equation}
with $a_1=-0.05$ and $a_2=-1$. The unactuated coordinate $x_1$ decays
slowly and drives the actuated coordinate $x_2$ through $x_1^2$.
Introducing $z_1=x_1$, $z_2=x_2$, and $z_3=x_1^2$ closes the dynamics
because
\[
 \dot z_1=a_1z_1,\;
 \dot z_2=a_2z_2-a_2z_3+u,\;
 \dot z_3=2x_1\dot x_1=2a_1z_3.
\]
Thus the lift $\Psi(x)=[\,x_1\ \ x_2\ \ x_1^2\,]^\top$ gives
\begin{equation}\label{eq:exactAB}
\begin{gathered}
 \dot z=Az+Bu,\qquad y=Cz,\\
 A=\begin{bmatrix}
 a_1&0&0\\
 0&a_2&-a_2\\
 0&0&2a_1
 \end{bmatrix},
 B=\begin{bmatrix}0\\1\\0\end{bmatrix},
 C=\begin{bmatrix}1&1&0\end{bmatrix}.
\end{gathered}
\end{equation}
Here
\[
 J_\Psi(x)=\begin{bmatrix}1&0\\0&1\\2x_1&0\end{bmatrix},
 \qquad g=\begin{bmatrix}0\\1\end{bmatrix},
\]
so $J_\Psi f=A\Psi$, $J_\Psi g=B$, and $h=C\Psi$ hold exactly.
The residuals $e_A,E_B,e_y$ from
Section~\ref{subsec:additive_compatibility} therefore vanish.
In particular, differentiating $x_1^2$ introduces no state-dependent
input term because $x_1$ is unactuated. Physical lifted states lie on
the two-dimensional set $z_3=z_1^2$ within $\R^3$.

The realization is observable because
$\det\col(C,CA,CA^2)=a_1a_2(a_1-a_2)\ne0$. In contrast,
$A^qB=a_2^qB$ gives
\begin{equation}\label{eq:exact_rank}
 \rank\mathcal C(A,B)=1<\rank J_\Psi=2<3=N.
\end{equation}
The uncontrollable eigenvalues $a_1,2a_1$ are stable, so the
realization is stabilizable. Also, $\ad_f^qg=(-a_2)^qg$ verifies
Proposition~\ref{prop:bracket_krylov}. Thus exact input compatibility
and observability do not require full controllability of the lifted
realization. Observability concerns distinguishing initial conditions
through their outputs, whereas controllability concerns changing the
state through the input. Here the unactuated coordinates affect the
output over time and are therefore observable, although the input cannot
change their autonomous evolution. Their different initial values across
trajectories can still provide information for identifying $A$; separate
identification of $A$ and $B$ depends on the regression rank condition in
Section~\ref{sec:integral}, rather than full lifted controllability.

\begin{figure*}[t]
\centering
\includegraphics[width=\textwidth]{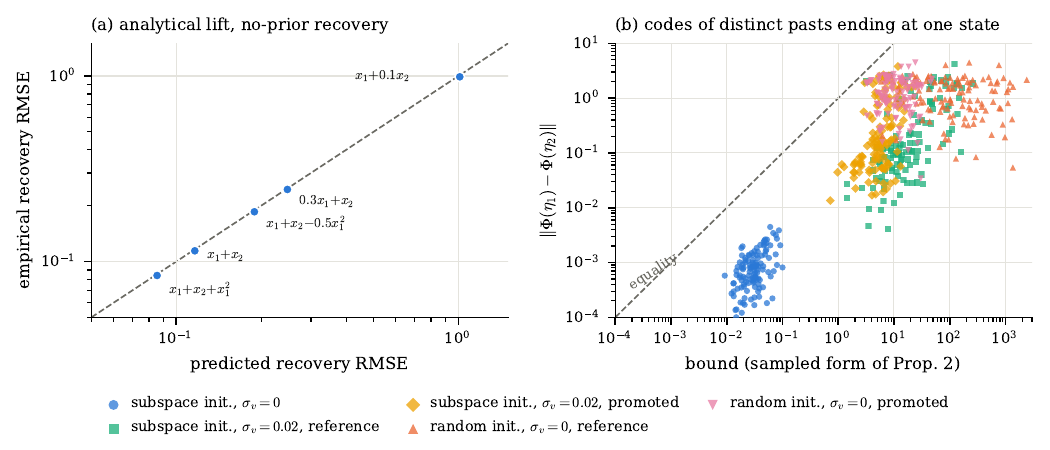}
\caption{Exact-system results.
(a) Predicted and empirical initial-state recovery errors for the
analytical lift with different observable output maps.
(b) Lifted-state differences for windows whose past trajectories
end at the same nonlinear state, compared with the sampled form
of~\eqref{eq:code_disagreement_bound}. Results from all three
identification runs are shown.}
\label{fig:s1}
\end{figure*}

\paragraph{Recovery with the analytical lift.}
We first assess initial-state recovery with the known matrices, so that
errors in the learned representation do not enter this comparison.
After subtracting the known-input response, the least-squares estimate
in Proposition~\ref{prop:posterior} has zero-mean error with covariance
$\varepsilon(W_{o,H}^{\Delta})^{-1}$. For a zero-mean error vector $e$,
$\mathbb E\norm{e}^2=\operatorname{tr}\operatorname{Cov}(e)$.
Consequently, the predicted root-mean-square Euclidean recovery error is
\begin{equation}\label{eq:oracle_rmse}
 E_{\rm rec}
 =
 \sqrt{
 \varepsilon\,
 \operatorname{tr}\bigl((W_{o,H}^{\Delta})^{-1}\bigr)
 },
 \qquad
 \varepsilon=\sigma_v^2\Delta t.
\end{equation}
Using exact covariance normalization from the identification
trajectories, $D_y=I$, $\Delta t=0.01$~s, $H=500$, and
$\sigma_v=0.02$ gives $E_{\rm rec}=0.1163$. The empirical value
over 200 noise realizations is $0.1142$, consistent with the nominal
covariance calculation. In terms of the Gramian eigenvalues,
$E_{\rm rec}^2=\varepsilon\sum_{i=1}^N\lambda_i(W_{o,H}^{\Delta})^{-1}$,
so weakly observable directions can dominate this Euclidean error.
This trace-based diagnostic is distinct from the log-determinant
objective, which measures the volume of nominal uncertainty.
Figure~\ref{fig:s1}(a) shows the corresponding comparison for
several observable output maps. These calculations concern
recovery of the normalized initial state of the known linear
model, without a prior or projection onto the physical lifted image.

\paragraph{Identification from input-output measurements.}
For learning, we use $N=3$, $\Delta t=0.1$~s, and both a measurement
window and prediction horizon of $5$~s. This sampling interval differs
from that used in the preceding analytical recovery experiment.
The window spans 50 input intervals and contains their 51 endpoint
outputs. Each data set contains 40 identification,
10 validation, and 20 test trajectories of $40$~s. Initial states
are uniform in $[-1.5,1.5]^2$. Inputs are random multilevel signals
with amplitudes uniform in $[0.3,1]$ and holding times uniform in
$[0.5,3]$~s. Three independent runs are performed for each setting.

Two additional diagnostics examine what the encoder has learned.
First, for evaluation only, a linear transformation $S\in\R^{3\times3}$
relating the identified coordinates to $\Psi(x)$ is fitted on validation
trajectories and then fixed. The relative discrepancy
$\norm{z_n-S\Psi(x)}/\norm{z_n}$ is evaluated on test windows without
measurement noise. Allowing $S$ accounts for the nonuniqueness of lifted
coordinates under a linear change of basis. A small discrepancy together
with nonsingular $S$ indicates agreement with the analytical lift on
these test data; it does not establish exact closure on an open set.

Second, we construct 40 pairs of windows without measurement noise whose
different past trajectories end at the same nonlinear state. A
state-based encoder would assign identical lifted states to each pair.
For each learned model, their initial difference $\delta z_0$ is
compared with the sampled counterpart of
\eqref{eq:code_disagreement_bound}, under a common future input.
Writing $e_{i,j}=D_y(y_j-C_n\hat z_{i,j})$ for the scaled prediction
error from window $i$, define
$E_i=(\Delta t\sum_{j=0}^{H-1}\norm{e_{i,j}}^2)^{1/2}$. For
$W_{o,H}^{\Delta}\succ0$, the same output-separation argument gives
\[
 \norm{\delta z_0}
 \le \frac{E_1+E_2}{\sqrt{\lambda_{\min}(W_{o,H}^{\Delta})}}
 \le \frac{2\max(E_1,E_2)}
 {\sqrt{\lambda_{\min}(W_{o,H}^{\Delta})}}.
\]
The common true output is well defined because the two trajectories
start from the same nonlinear state and receive the same future input.
The sampled sums use the same output scaling and sample indices as the
Gramian.

\begin{table*}[t]
\centering
\small
\caption{Exact-system results, averaged over three identification
runs. Prediction RMSE uses measured test outputs over $5$~s.
$W=W_{o,H}^{\Delta}$ uses the normalization and output scaling of
Section~\ref{sec:learning}. State difference and bound are averages
of the medians over the 40 paired windows without measurement noise.
``Optimized'' denotes the result of~\eqref{eq:obsloss}.
Values are rounded to the displayed precision.}
\label{tab:e1}
\setlength{\tabcolsep}{4pt}
\begin{tabular}{@{}lcccc@{}}
\toprule
Model
& Prediction RMSE
& $\lambda_{\min}(W)$
& State difference
& Bound\\
\midrule
\multicolumn{5}{@{}l}{
\emph{Identification without measurement noise}}\\
Analytical lift
& $0.000$
& $4.50\cdot10^{-4}$
& ---
& ---\\
Subspace, reference/optimized
& $0.000$
& $4.63\cdot10^{-4}$
& $8.0\cdot10^{-4}$
& $0.031$\\
Random, reference
& $0.016$
& $1.65\cdot10^{-6}$
& $1.3$
& $240$\\
Random, optimized
& $0.016$
& $8.86\cdot10^{-5}$
& $1.3$
& $11$\\
\midrule
\multicolumn{5}{@{}l}{
\emph{Identification with $\sigma_v=0.02$}}\\
Analytical lift
& $0.020$
& $4.49\cdot10^{-4}$
& ---
& ---\\
Subspace, reference
& $0.027$
& $5.11\cdot10^{-5}$
& $0.52$
& $25$\\
Subspace, optimized
& $0.028$
& $4.96\cdot10^{-4}$
& $0.45$
& $5.9$\\
\bottomrule
\end{tabular}
\end{table*}

Without measurement noise, subspace initialization
\citep{vanoverschee1996subspace} gives a representation close to
the analytical lift: the relative error after the change of
coordinates is below $4\cdot10^{-4}$, the fitted transformation
$S$ is nonsingular, and the controllability matrix is numerically
of rank one ($\sigma_2/\sigma_1<10^{-5}$).
The observability optimization finds no admissible improvement
and retains the reference in all three runs. This shows that the
learning procedure can recover the analytical representation closely
under this initialization and data setting. The prediction errors displayed
as $0.000$ in Table~\ref{tab:e1} are rounded values.

Random initialization gives prediction RMSE $0.016$, but lifted
states associated with the same current nonlinear state differ
by about $1.3$ (Table~\ref{tab:e1}).
For each of the six reference and optimized models, the median
fraction of squared difference along the least observable
direction exceeds $99.5\%$. If $q_{\min}$ is a unit eigenvector
associated with the smallest Gramian eigenvalue, this fraction is
$(q_{\min}^\top\delta z_0)^2/\norm{\delta z_0}^2$.
Thus most of the history-dependent difference lies in a direction
with weak output sensitivity. Optimizing observability increases
the mean smallest Gramian eigenvalue by about a factor of 50
and reduces the reported bound from about 240 to 11, while
leaving the actual state differences essentially unchanged
(Figure~\ref{fig:s1}(b)). Accurate output prediction therefore
does not by itself establish a state-based representation.
There is no conflict with Proposition~\ref{prop:code_disagreement}:
the bound limits the possible discrepancy, but need not be tight, and
the reported bound of 11 remains well above the discrepancy of 1.3.
Moreover, the table reports averages of pairwise medians, rather than
a worst-case guarantee over all histories.

With measurement noise, observability optimization brings the
smallest Gramian eigenvalue close to that of the analytical lift,
with a small increase in prediction error. The average state
difference decreases from $0.52$ to $0.45$, but the relative
error after the change of coordinates remains near $0.6$.
The change in the Gramian is therefore much larger than the change in
the same-state discrepancy, and agreement with the analytical coordinates
remains limited. Prediction RMSE in this table is measured against noisy
test outputs, which explains the value $0.020$ even for the analytical
model. These results distinguish three properties: output accuracy,
nominal observability, and recovery of a state-based lifting. Improvement
in the second does not by itself establish the third.

\subsection{Controlled Duffing oscillator}
\label{subsec:S2}

\subsubsection{System and identification data}
\label{sec:duffing_analytic}
\label{sec:duffing_setup}

Consider the controlled Duffing system
\citep{korda2018transient,korda2020optimal}:
\begin{equation}\label{eq:duffing}
\begin{aligned}
 \dot x_1&=x_2,\\
 \dot x_2&=-0.5x_2+x_1-4x_1^3+0.5u,\\
 y^{\rm m}&=x_1+v.
\end{aligned}
\end{equation}
For $u=0$, equilibrium requires $x_2=0$ and
$x_1-4x_1^3=0$, giving $(0,0)$ and $(\pm0.5,0)$.
The autonomous Jacobian is
\[
 J_f(x)=\begin{bmatrix}0&1\\1-12x_1^2&-0.5\end{bmatrix}.
\]
At $(\pm0.5,0)$ it has negative trace and positive determinant, so
both equilibria are asymptotically stable. At the origin its determinant
is negative, giving a saddle. The ideal output also satisfies
$\col(h,\mathcal L_fh)=\col(x_1,x_2)$, whose Jacobian is the identity.
Thus the nonlinear state is locally differentially observable from
position, even though only position is measured. The following result shows that observability of the nonlinear
system does not ensure existence of an informative exact additive lift.

\begin{proposition}[Exact additive lifting of the controlled Duffing system]
\label{prop:duffing_nolift}
Let $\Psi$ be $C^1$ on a nonempty open set
$\mathcal O\subset\R^2$ and assume it satisfies~\eqref{eq:exact_conditions}
for~\eqref{eq:duffing}, with constant $A,B$.
Then $B=0$ and $\Psi$ is constant on each connected component
of $\mathcal O$, with $A\Psi=0$.
\end{proposition}

\begin{proof}
On any open rectangle $I_1\times I_2\subset\mathcal O$,
the input vector field is $g=[\,0\ \ 0.5\,]^\top$.
Consequently, $J_\Psi g=B$ gives
$\partial\Psi/\partial x_2=2B=:b$, where $b$ is constant, and
$\Psi(x)=\Gamma(x_1)+bx_2$.
Substituting this form in $J_\Psi f=A\Psi$ gives
\[
 \bigl(\Gamma'(x_1)-0.5b\bigr)x_2+b(x_1-4x_1^3)
 =A\Gamma(x_1)+Abx_2.
\]
Since this identity holds throughout the rectangle, the coefficient of
$x_2$ and the remaining terms must agree separately:
\[
 \Gamma'(x_1)=(A+0.5I)b,
 \qquad
 A\Gamma(x_1)=b(x_1-4x_1^3).
\]
The first equality makes $\Gamma$ affine in $x_1$, so $A\Gamma$ is
also affine. The cubic coefficient in the second equality therefore
forces $b=0$, and the first then gives
$\Gamma'=0$. Thus $\Psi$ is locally constant throughout
$\mathcal O$, and hence constant on each connected component,
with $A\Psi=0$.
\end{proof}

Since $h(x)=x_1$ is nonconstant, such a lift cannot also satisfy
$h=C\Psi$. Together with
Proposition~\ref{prop:encoder_factorization}, this excludes an
exact observable window-based realization under its trajectory
assumptions. The obstruction concerns simultaneous exact autonomous
closure and a constant additive input matrix for the original input
$u$; increasing the finite lifted dimension does not remove it.
We therefore assess an approximate additive model, for which prediction,
state estimation, and control must be evaluated empirically. The
proposition does not rule out accurate approximation on the sampled
operating region.

We use $\Delta t=0.01$~s, a $0.5$~s measurement window, a
$1$~s prediction horizon, sensor-noise standard deviation
$\sigma_v=0.02$, and $|u|\le1$.
Each identification set contains 60 trajectories of $5.5$~s
from initial states uniform in $[-1,1]^2$: 40 use random
multilevel inputs with amplitudes in $[0.3,1]$ and holding
times in $[0.1,1]$~s; 20 use saturated state feedback with
additional excitation of amplitude $0.3$, towards set-points
in $[-0.6,0.6]$.
The random inputs provide varied forced responses, while feedback
provides coverage near the saddle and intermediate operating points that
uncontrolled trajectories need not visit for long. The additional
excitation introduces input variation beyond the feedback action, as
motivated in Section~\ref{sec:integral}. Full-state feedback is used
only to generate part of this simulated data set; identification receives
only the resulting inputs and noisy outputs.
There are 15 validation trajectories of the same design and
30 open-loop test trajectories, ten at each input amplitude
$0.25$, $0.5$, and $1$. These test amplitudes examine the additive model
under different input strengths, including the contribution of input-map
mismatch described by~\eqref{eq:residual_error_bound}.
Trajectories are assigned to identification, validation, and test sets
before windows are formed, preventing overlapping windows from one
trajectory from being shared across those sets.

Five independent identification runs use different identification
and validation data and a common test set. Thus variation across runs
reflects different identification and validation records, while comparisons
within each run use the same evaluation trajectories. The dimension is fixed
at $N=6$, selected on the first run as the smallest value in
$\{4,6,8\}$ whose validation prediction error is within $5\%$
of the best.
Each reference model is compared with the result of
\eqref{eq:obsloss}, using $\delta_p=\delta_c=0.1$ and the same
data, dimension, normalization, observer-tuning procedure, and
controller, so only the constrained observability stage differs. One run
returns the reference because no admissible improvement is found; it
remains in all averages.

A fixed-basis predictor provides an additional comparison
within the lifted-model approach of \citet{korda2018linear}.
It uses six output delays spaced by $0.1$~s, five input
averages over $0.1$~s intervals, and 40 radial basis functions,
giving $N=51$. It is identified by the same normalization
and integral regression. Because its basis and dimension differ, only its
prediction and control errors are compared; its Gramian determinant is not
included in the $N=6$ score comparison.

\subsubsection{Prediction and state estimation}
\label{sec:duffing_results}

The integral regression has $\chi_I=0.015$--$0.042$, corresponding to
spectral condition numbers of approximately 24--67. The nonzero ratios
indicate full row rank of the reported regressors, although some directions
are less strongly represented than others.
All selected optimized models satisfy the two identification-error
bounds, with maximum ratios $\Lpred/b_p=0.9853$ and
$\Lcons/b_c=0.9827$; generalization is assessed separately on the test
trajectories.
Table~\ref{tab:duffing_results} and Figure~\ref{fig:duffing}(a)
compare predictions from the window-based estimate and the
recursive observer, using identical starting times, future
inputs, and prediction steps after a $1$~s observer
initialization period. At each selected starting time, the two estimators
supply alternative initial lifted states to the same model. Subsequent
predictions are propagated without measurement corrections. Hence this
comparison measures the effect of initialization on future-output
prediction, using the same future inputs and prediction horizon.
The test RMSE excludes the initial output and uses the 100 subsequent
samples. It is evaluated against outputs without measurement noise,
whereas the training loss uses measured outputs.

\begin{table*}[t]
\centering
\small
\caption{Duffing output prediction over $1$~s: means $\pm$
descriptive standard deviations over five identification runs.
Errors use the same starting times and prediction steps 1--100,
relative to outputs without measurement noise.
$W=W_{o,H}^{\Delta}$ is in the regularized normalized coordinates;
its determinant is unregularized.}
\label{tab:duffing_results}
\setlength{\tabcolsep}{5pt}
\begin{tabular}{@{}lccc@{}}
\toprule
Model
& $\log\det W$
& \shortstack{Window-initialized\\prediction RMSE}
& \shortstack{Observer-initialized\\prediction RMSE}\\
\midrule
Reference
& $-42.2\pm2.2$
& $0.018\pm0.002$
& $0.064\pm0.001$\\
Optimized
& $-33.3\pm6.4$
& $0.018\pm0.002$
& $0.066\pm0.002$\\
Fixed basis ($N=51$)
& ---
& $0.041\pm0.002$
& $0.081\pm0.006$\\
\bottomrule
\end{tabular}
\end{table*}

The nominal Gramian criterion improves, while window-initialized
prediction RMSE remains $0.018$ to the displayed precision. The table
reports $\log\det W$, with no determinant regularization, whereas
the optimized score is $-\ell_{\rm obs}=\log\det(W+\varepsilon I_N)$.

The magnitude of the score improvement depends on normalization: only
three of the six covariance eigenvalues exceed $\beta$. Regularization
therefore materially affects the scaling of several lifted directions.
Re-expressing each fixed predictor using exact covariance
normalization preserves its input-output behaviour.
In these coordinates, the regularized score $-\ell_{\rm obs}$
still increases in all four changed pairs, but by $1.04$--$2.82$,
compared with $6.31$--$13.98$ under the reported regularized
normalization. The improvement therefore persists, but its size
is sensitive to the coordinate convention.

The recursive observers are stable, with spectral radii of
$(I-L_\infty C_n)A_d$ between $0.94$ and $0.98$.
Nevertheless, their estimates give larger future-output
prediction errors than the window-based estimates on the same
models. Observability optimization slightly increases
observer-initialized prediction error in all four changed pairs.
The homogeneous nominal observer error therefore decays, but continuing
model mismatch and errors in the encoded coordinates still affect the
estimate.

Figure~\ref{fig:duffing}(b) also evaluates physical-state estimates using
linear readout maps fitted on validation trajectories. Those maps are
used only for evaluation and are not supplied to the controller or used
to train the lifted representation. Taken together, the matched
future-output predictions and the observer stability check show that
better nominal Gramian values do not resolve the estimation limitation
in this experiment.

\subsubsection{Output-feedback tracking}
\label{sec:duffing_control}

The reference sequence is $r=0.5,-0.5,0,0.25$, with $5$~s
at each value. This tests transitions between the stable
equilibria, holding the saddle, and tracking a point requiring
nonzero equilibrium input. Holding $y=r$ requires $x_1=r$ and
$x_2=0$. Substitution into the second state equation gives
$0=r-4r^3+0.5u$, so the true equilibrium input is
\begin{equation}\label{eq:duffing_target_input}
 u_{\rm eq}(r)=8r^3-2r.
\end{equation}
The values $r=\pm0.5$ and $r=0$ require zero equilibrium input,
while $u_{\rm eq}(0.25)=-0.375$. Thus all four references admit a
true equilibrium under $|u|\le1$, although holding the saddle requires
stabilizing feedback.
The identified controllers use~\eqref{eq:target}; they are not
supplied with~\eqref{eq:duffing_target_input}.

Each fitted model is evaluated from 20 initial states uniform
in $[-0.8,0.8]\times[-0.5,0.5]$, with the same measurement-noise
realizations across controllers.
A $0.5$~s interval with $u=0$ supplies the initial measurement
window. We also compare the linearization at the origin,
\[
 A_{\rm lin}=\begin{bmatrix}0&1\\1&-0.5\end{bmatrix},\;
 B_{\rm lin}=\begin{bmatrix}0\\0.5\end{bmatrix},\;
 C_{\rm lin}=\begin{bmatrix}1&0\end{bmatrix},
\]
with a Kalman filter and the same MPC, and the true nonlinear model
with an extended Kalman filter and nonlinear MPC
\citep{diehl2005rti}.
The latter uses the true equilibrium input; all controllers
use the same horizon, weights, and input bounds. The local linear model
tests whether linearization at one operating point is adequate for this
task. The true-model controller provides a comparison with access to the
nonlinear dynamics and equilibrium-input relation; its numerical solution
is not a certified performance lower bound for the learned controllers.

Let $e_{ik}=x_{1,ik}-r_k$ and let $\mathcal I_j$ contain the
100 samples in the final second of reference segment $j$.
Define
\[
 b_{ij}=\frac1{100}\sum_{k\in\mathcal I_j}e_{ik},
 \;
 B_i=\frac14\sum_{j=1}^4|b_{ij}|.
\]
Here $b_{ij}$ is the signed mean error of trial $i$ in reference segment
$j$. Taking its absolute value before averaging prevents positive and
negative offsets in different segments from cancelling. The reported
bias averages $B_i$ over trials. Final-second RMSE is computed from
squared errors over these same windows. The distinction is visible in
the identity
\[
 \frac1{100}\sum_{k\in\mathcal I_j}e_{ik}^2
 =b_{ij}^2+
 \frac1{100}\sum_{k\in\mathcal I_j}(e_{ik}-b_{ij})^2.
\]
Thus RMSE includes both the segment offset and fluctuations about it.
Using the final second reduces the influence of transitions, but these
finite-window measures do not establish asymptotic steady-state errors.
A trial fails the tracking test if
$|b_{ij}|>0.1$ for any segment.
All 640 closed-loop trials are included: 600 trials for the three
identified model types with two estimators over five runs, and 20 each
for the local linear and true-model controllers. All remain finite,
satisfy the input bound, and have
$\max_k\norm{x_{ik}}_\infty<5$.
Nominal-target infeasibility is recorded separately.

\begin{table*}[t]
\centering
\small
\caption{Duffing tracking results.
``Observer'' uses~\eqref{eq:observer}; ``Window''
uses~\eqref{eq:window_estimate} at every sample.
Means $\pm$ descriptive standard deviations are over five
identification runs, each with 20 trials.
The local linear and true-model controllers use 20 trials each.
The final column counts trials containing an infeasible
nominal steady target.}
\label{tab:duffing_cl}
\setlength{\tabcolsep}{3.5pt}
\begin{tabular}{@{}llcccc@{}}
\toprule
Model
& Estimator
& \shortstack{Final-second\\RMSE}
& Bias
& \shortstack{Tracking\\failures}
& \shortstack{Infeasible\\targets}\\
\midrule
Reference & Observer
& $0.030\pm0.007$
& $0.020\pm0.006$
& 1/100 & 0/100\\
Optimized & Observer
& $0.034\pm0.009$
& $0.023\pm0.007$
& 0/100 & 20/100\\
Reference & Window
& $0.010\pm0.002$
& $0.005\pm0.002$
& 0/100 & 0/100\\
Optimized & Window
& $0.011\pm0.003$
& $0.007\pm0.004$
& 0/100 & 20/100\\
Fixed basis & Observer
& $0.061\pm0.022$
& $0.046\pm0.016$
& 29/100 & 0/100\\
Fixed basis & Window
& $0.023\pm0.002$
& $0.017\pm0.002$
& 0/100 & 0/100\\
Local linear & Kalman filter
& $0.112$
& $0.086$
& 20/20 & 0/20\\
True model & Extended KF
& $0.009$
& $0.004$
& 0/20 & 0/20\\
\bottomrule
\end{tabular}
\end{table*}

\begin{figure*}[t]
\centering
\includegraphics[width=\textwidth]{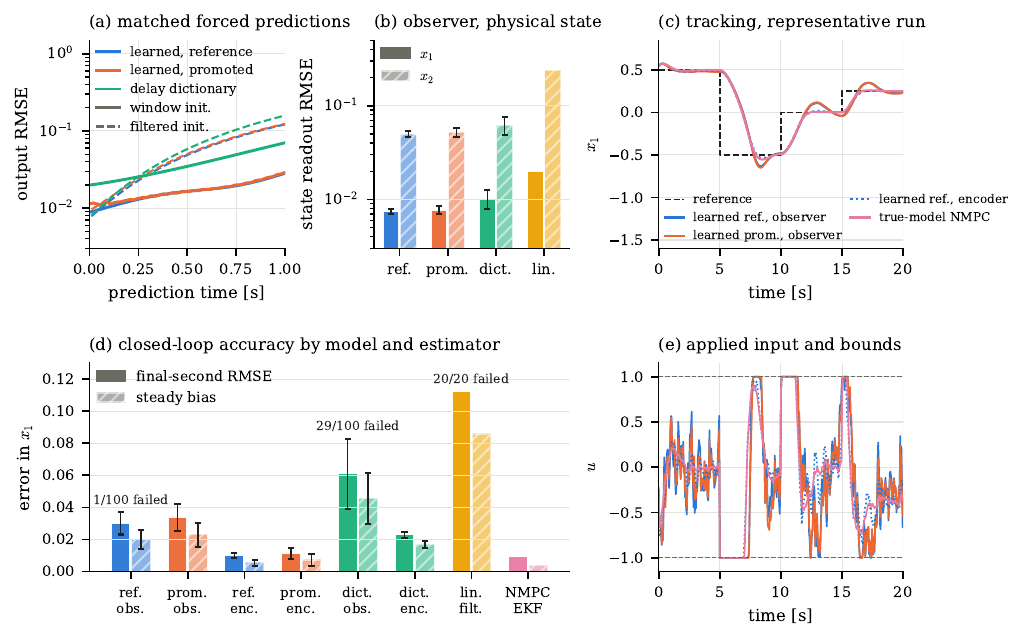}
\caption{Duffing results.
(a) Output-prediction errors from window-based estimates (solid)
and recursive-observer estimates (dashed), using identical
starting times and future inputs.
(b) Physical-state estimation errors, evaluated using linear
maps fitted on validation trajectories for evaluation only.
(c) Output tracking and (e) applied inputs for the same
closed-loop trial.
(d) Final-second tracking RMSE and bias, with tracking failures
indicated; bars show means and descriptive standard deviations
over identification runs.
The labels ``promoted'' and ``dictionary'' refer to the optimized
and fixed-basis models, respectively.
Nominal-target infeasibility is reported separately in
Table~\ref{tab:duffing_cl}.}
\label{fig:duffing}
\end{figure*}

Table~\ref{tab:duffing_cl} and Figure~\ref{fig:duffing} show that
observability optimization does not improve mean final-second
tracking error or bias with the recursive observer.
Both errors increase in each of the four changed pairs,
although the threshold-based failure count decreases from one
trial to zero; removing a single threshold violation can coexist with
larger average errors, so this change is not evidence of improved mean
tracking accuracy.
Recomputing the estimate from the measurement window has a
larger effect: for the reference model, final-second RMSE
decreases from $0.030$ to $0.010$ and bias from $0.020$ to
$0.005$. The fixed-basis model also improves, while the local
linear controller fails the tracking test in every trial. Since the
identified model and controller are unchanged, within this protocol the
estimator choice has a larger effect than the observability-optimization
stage; this need not hold for other systems, noise levels, or model
classes.

The true-model controller achieves final-second RMSE $0.009$.
Its similar aggregate accuracy to the reference model with
window-based estimation does not establish controller
equivalence, since it uses the known nonlinear dynamics
and equilibrium input.

The optimized model from the first identification run requires
$|u_s|=1.3769$ at $r=\pm0.5$, violating $|u|\le1$. For this model, $\kappa(\mathcal{T})=2.9\cdot10^{5}$ and $G(1)=-0.363$, so $\mathcal{T}$
is nonsingular and~\eqref{eq:target_unique} applies.
By~\eqref{eq:target_unique}, all 20 trials for this model therefore contain
infeasible nominal targets, even though the tracking test is passed.
Clipping produces an admissible input-cost centre but does
not restore the target equations. The actual nonlinear equilibria at
$r=\pm0.5$ require $u=0$, so the reported nominal infeasibility reflects
the identified model's steady-target relation. It does not indicate
infeasibility of the corresponding true equilibria. Nor does it make the
input-only MPC problem infeasible: the target equalities are not hard
constraints in~\eqref{eq:mpc}, and the clipped value enters only its
input penalty. This explains how a trial can pass the tracking test
while containing an infeasible nominal steady target. There is also a structural limitation in the nominal target calculation.
With $\mathcal T$ as in Section~\ref{sec:ofc}, the minimum-norm
least-squares target before clipping is
$\col(z_s,u_s)=\mathcal T^\dagger\col(0,r)$, where
$\mathcal T^\dagger$ denotes the Moore--Penrose inverse. For fixed
matrices and scalar input and output, this gives $u_s=\gamma r$ for
some constant $\gamma$. It cannot reproduce the cubic relation
\eqref{eq:duffing_target_input} over an interval. Good short-horizon
predictions therefore need not imply an accurate nominal equilibrium-input
relation, and neither the observability objective nor the error
constraints enforce it.
\section{Conclusions}
\label{sec:conclusions}

This paper has developed a continuous-time framework for
identifying control-relevant Koopman representations from
input-output trajectories without full-state measurements.
The method combines integral identification with finite-horizon
observability optimization subject to separate bounds on
output-prediction and lifted-state consistency errors.
Coordinate normalization controls the scale of the observability
criterion, which has an estimation interpretation under the
stated nominal Gaussian assumptions.

The analysis distinguishes observability, input-map compatibility,
controllability, and data informativity.
Under the stated exactness and observability assumptions,
the representation constructed from an input-output window
depends only on the current nonlinear state.
For approximate representations, prediction errors and the
smallest observability Gramian eigenvalue bound the differences
between lifted states associated with the same nonlinear state.
Exact additive lifting also imposes a controllability restriction:
when $N>n$, the lifted realization cannot be fully controllable.
Controllability is therefore assessed as a structural property,
while observability is optimized subject to the accuracy
constraints.

The numerical examples show that improved nominal observability
does not by itself ensure better estimation or control.
For the exactly liftable system, subspace initialization recovers
a representation close to the analytical lift in the absence of
measurement noise. Random initialization can instead retain
dependence on past trajectories despite accurate output
prediction.
For Duffing, the observability criterion improves without
improving observer-based prediction or mean final-second
tracking accuracy. Recomputing the lifted-state estimate from
each measurement window has a larger effect on tracking.
The Gramian improvement persists under exact covariance
normalization, although its magnitude depends on the
normalization convention. Infeasible nominal steady targets
further show that successful tracking does not establish exact
equilibrium compatibility.

These findings motivate observability criteria that account
for model error and the implemented estimator, together with
output-feedback controllers based on certified error bounds.
The present empirical constraints and nominal controller do
not provide nonlinear closed-loop stability or offset-free
tracking guarantees.

\appendix
\section{Implementation details}
\label{app:implementation}

\subsection{Representation and initialization}
\label{app:encoder}
The window map uses
\begin{equation*}
\begin{aligned}
&F_\theta(\eta)=L_\theta\eta+
 W_3\tanh\!\bigl(W_2\tanh(W_1\eta+b_1)\\
 & +b_2\bigr) +b_3,\\
 &\Phi_\theta(\eta)=F_\theta(\eta)-F_\theta(0),
\end{aligned}    
\end{equation*}
with two intermediate vectors of dimension 64. Window outputs are scaled
by $D_y$ in~\eqref{eq:Dy} and inputs by $1/u_{\max}$. 

For subspace initialization, the
future outputs are regressed on the window and future inputs with
regularization $10^{-6}$. If $G$ is the window coefficient and $S_p$
the regularized uncentered second moment of the past window, the
factorization
$GS_p^{1/2}=U\Sigma V^\top$ gives
$L_\theta=\Sigma_N^{1/2}V_N^\top S_p^{-1/2}$.
The nonlinear output matrix is initialized with standard deviation
$0.03/\sqrt{64}$. The fixed-basis comparison uses thin-plate radial
basis functions centred at randomly selected identification samples,
with the basis value at the origin subtracted.

\subsection{Model identification and constrained optimization}
\label{app:optimization}
The regularization parameters are
\[\lambda_{AB}=10^{-6}\Delta t^2,\;\lambda_C=10^{-6},\; \beta=10^{-3}\operatorname{tr}(\Sigma_z)/N\]
where $\beta$ keeps the
normalization~\eqref{eq:normalized} well conditioned, and
$\lambda_{AB},\lambda_C$ regularize the regressions~\eqref{eq:AB_ls}.

For the scalar-output experiments,
$\varepsilon=(D_y\sigma_v)^2\Delta t$, corresponding to
$\sigma_s=D_y\sigma_v$ in Proposition~\ref{prop:posterior}, using $\sigma_v=0.02$ also
when identification data contain no measurement noise. In that case
$\varepsilon$ is purely numerical regularization.
We use 
\[\lambda_c=1,\;\delta_p=\delta_c=0.1,\;\tau_p=\tau_c=10^{-6}\] with no additional constraint-acceptance
tolerance. Prediction segments start every five samples for the exact
system and every 20 samples for Duffing.

Both stages use Adam \citep{kingma2015adam}. The reference fit uses
an initial step size $2\cdot10^{-3}$ with cosine decay, gradient-norm
clipping at 10, and validation every 50 iterations. The constrained
stage uses an inequality augmented Lagrangian
\citep{nocedal2006numerical}, step size $5\cdot10^{-4}$, penalty 50,
and multiplier updates every 25 iterations. The iteration limits for
the two stages are 1500/1000 for the exact system with subspace
initialization, 3000/1500 with random initialization, and 2500/1500
for Duffing.

\subsection{State estimation and predictive control}
\label{app:control}

The observer parameter is selected from
$q\in\{0.01,0.03,0.1,0.3,1,3,10,30,100\}$ by $1$~s output prediction
after a $1$~s initialization period on validation trajectories.
The local linear observer uses $\widehat Q=\Delta t I_2$.
The MPC settings are \[H_c=100,\;Q=1,\; Q_f=10,\; R=0.01,\; R_\Delta=0.1.\] Each quadratic program uses 300 accelerated
projected-gradient iterations \citep{beck2009fista}, initialized from
the shifted previous solution. The true-model nonlinear MPC uses two
Gauss--Newton iterations per sample, with the same quadratic-program
solver and an extended Kalman filter with process covariance
$10^{-6}I_2$.

Simulations use double precision and fourth-order Runge--Kutta integration
with ten substeps per sample for the exact system and four for Duffing.
Reported standard deviations across the five identification runs use
divisor five. The accompanying code specifies the data generators,
random seeds, and selected solutions; no test results are used for
model selection or observer tuning.

\IfFileExists{elsarticle-harv.bst}
 {\bibliographystyle{elsarticle-harv}}
 {\bibliographystyle{plainnat}}

{\footnotesize
\setlength{\bibsep}{0pt}
\bibliography{bibfile}
}
\end{document}